\documentclass[12pt,epsf,amssymb]{article} \textheight =23 truecm
\def\lsim{\raise0.3ex\hbox{$<$\kern-0.75em\raise-1.1ex\hbox{$\sim$}}}
\def\gsim{\raise0.3ex\hbox{$>$\kern-0.75em\raise-1.1ex\hbox{$\sim$}}}
\def\noi{\noindent}  \def\bea{\begin{eqnarray}}
\def\eea{\end{eqnarray}} \def\beq{\begin{equation}}
\def\eeq{\end{equation}} 
\def\beeq{\begin{eqnarray}} \def\eeeq{\end{eqnarray}} \def\R{ {\rm R
\kern -.31cm I \kern .15cm}} \def\C{ {\rm C \kern -.15cm \vrule
width.5pt \kern .12cm}} \def\Z{ {\rm Z \kern -.27cm \angle \kern
.02cm}} \def\N{ {\rm N \kern -.26cm \vrule width.4pt \kern .10cm}}
\def\1{{\rm 1\mskip-4.5mu l} }

\usepackage{graphics} \usepackage{graphicx} \usepackage{epsfig}

\begin{document} \begin{center} 

{\large \bf Isgur-Wise functions and unitary representations \\ of the Lorentz group : the baryon case j = 0} 
\par \vskip 1 truecm

 {\bf A. Le Yaouanc, L. Oliver and J.-C. Raynal}
\par \vskip 1 truemm

{\it Laboratoire de Physique Th\'eorique}\footnote{Unit\'e Mixte de
Recherche UMR 8627 - CNRS }\\    {\it Universit\'e de Paris XI,
B\^atiment 210, 91405 Orsay Cedex, France} 

\end{center}
\par \vskip 3 truemm

\begin{abstract} 
We propose a group theoretical method to study Isgur-Wise (IW) functions. A current matrix element splits into a heavy quark matrix element and an overlap of the initial and final clouds, related to the IW functions, that contain the long distance physics. The light cloud belongs to the Hilbert space of a unitary representation of the Lorentz group. Decomposing into irreducible representations one obtains the IW function as an integral formula, superposition of {\it irreducible} IW functions with {\it positive measures}, providing positivity bounds on its derivatives. Our method is equivalent to the sum rule approach, but sheds another light on the physics and summarizes and gives all its possible constraints. We expose the general formalism, thoroughly applying it to the case j = 0 for the light cloud, relevant to the semileptonic decay $\Lambda_b \to \Lambda_c \ell \overline{\nu}_{\ell}$. In this case, the principal series of the representations contribute, and also the supplementary series. We recover the bound for the curvature of the $j = 0$  IW function $\xi_{\Lambda} (w)$ that we did obtain from the sum rule method, and we get new bounds for higher derivatives. We demonstrate also that if the lower bound for the curvature is saturated, then $\xi_{\Lambda} (w)$ is completely determined, given by an explicit elementary function. We give criteria to decide if any ansatz for the Isgur-Wise function is compatible or not with the sum rules. We apply the method to some simple model forms proposed in the literature. Dealing with a Hilbert space, the sum rules are {\it convergent}, but this feature does not survive hard gluon radiative corrections. 

\vskip 5 truemm
\noi LPT Orsay 09-08 \qquad\qquad March 2009 \par 
\vskip 8 truemm

\end{abstract}

\tableofcontents

\vskip 0.3 truecm

\section{Introduction} \hspace*{\parindent} 

The heavy quark limit of QCD and, more generally, Heavy Quark Effective Theory, has aroused an enormous interest in the decade of the 1990's, starting from the formulation of Heavy Quark Symmetry by Isgur and Wise \cite{1r}.\par
Then, the theoretical study of the properties of this limit did slow down, due essentially to the fact that Flavor Physics had more urgent domains to explore, like the determination of $|V_{ub}|$, the study of rare decays like $B \to X_s \gamma$, and the comparison of the many CP violation observables with the predictions of the Standard Model. Presently, the main interest is focussed on the search of New Physics, in view of the possibilities of the future experimental projects : LHCb, Super-Belle and the Super B Factory.\par
The present paper shows that the richness of the heavy quark limit of QCD had not been explored in the past in all its depth. The method proposed here allows to obtain important constraints on the Isgur-Wise (IW) functions, that carry the long distance QCD physics in the heavy quark limit. These constraints turn out to have simple and explicit phenomenological applications that can be tested at present or in the future.\par
As it is well-known, in the heavy quark limit, QCD possesses new global symmetries, namely the spin-flavor symmetry $SU(2N_f)$ where $N_f$ is the number of heavy quark flavors, in practice the b and c quarks.\par
Hadrons with one heavy quark such that $m_Q >> \Lambda_{QCD}$ can be thought as a bound state of a light cloud in the color source of the heavy quark. Due to its heavy mass, the latter is unaffected by the interaction with soft gluons.\par
In this approximation, the decay of a heavy hadron with four-velocity $v$ into another hadron with velocity $v'$, for example the semileptonic decay $\overline{B} \to D^{(*)}\ell \overline{\nu}_{\ell}$ or $\Lambda_b \to \Lambda_c \ell \overline{\nu}_{\ell}$, occurs just by free heavy quark decay produced by a current, and the rearrangement of the light cloud or "brown muck", to follow the heavy quark in the final state and constitute the final heavy hadron.\par
The dynamics is contained in the complicated light cloud, that concerns long distance QCD and is not calculable from first principles. Therefore, one needs to parametrize this physics through form factors, the IW functions.\par
The matrix element of a current between heavy hadrons containing heavy quarks $Q$ and $Q'$ can thus be factorized as follows \cite{2r}

$$<H'(v')|J^{Q'Q}(q)|H(v)>\ =\ <Q'(v'),\pm{1\over2}|J^{Q'Q}(q)|Q(v),\pm{1\over2}>$$
\beq
\label{1e}
<light,v',j',M'|light,v,j,M> 
\eeq
where $v$, $v'$ are the initial and final four-velocities, and $j$, $j'$, $M$, $M'$ are the angular momenta and corresponding projections of the initial and final light clouds. \par
The current affects only the heavy quark, and all the soft dynamics is contained in the {\it overlap} between the initial and final light clouds $<v',j',M'|v,j,M>$, that follow the heavy quarks with the same four-velocity. From now on it would be understood that this scalar product concerns the light cloud. This overlap is independent of the current heavy quark matrix element, but depends on the four-velocities $v$ and $v'$. The IW functions are precisely given by these light clouds overlaps.\par
An important hypothesis has been done in writing the previous expression, namely neglecting {\it hard gluon radiative corrections}, that we will assume from now on.\par
As we will make explicit below, the light cloud belongs to a Hilbert space, and transforms according to a unitary representation of the Lorentz group. Then, as we will show, the whole problem of getting rigorous constraints on the IW functions amounts to decompose unitary representations of the Lorentz group into irreducible ones. This will allow to obtain for the IW functions general integral formulas in which the crucial point is that {\it  the measures are positive}.\par
In \cite{3r}, for bound states made up of a heavy quark and a non-relativistic light quark, we did already exploit the {\it positivity} of matrices of moments of the ground state wave function, that allowed to bound the derivatives of the corresponding IW function $\xi_{NR}(w)$, where $NR$ stands for the non-relativistic approximation for the light quark. The present paper extends this method to the non-trivial case of the true QCD in the heavy quark limit.\par
We treat here the case of a light cloud with angular momentum $j = 0$ in the initial and final states, as happens in the baryon semileptonic decay $\Lambda_b \to \Lambda_c \ell \overline{\nu}_{\ell}$ where, in the quark model, the light diquark system has $S = 0$ with orbital angular momentum $L = 0$ relative to the heavy quark, and therefore $j = 0$ in the relativistic language. The whole spin of the baryon is carried by the heavy active quark.\par
A different but, as we will show below, equivalent method to the one of the present paper was developed in a number of articles using {\it sum rules} in the heavy quark limit, like the famous Bjorken sum rule and its generalizations \cite{4r}\cite{5r}\cite{6r}\cite{7r}.\par
Although in our previous papers and also in most work by other authors, the sum rules are formulated using the heavy {\it hadron} states, they could be formulated in an equivalent way using only the light cloud, the reason being that the heavy quark spin decouples from the soft QCD physics.\par
The sum rule method is completely equivalent to the method of the present paper. Indeed, starting from the sum rules one can demonstrate that an IW function, say $\xi(v.v') =\ <v'|v>$ in a simplified notation, is a function of {\it positive type}, and that one can construct a unitary representation of the Lorentz group $U(\Lambda)$ and a vector state $|\phi_0>$ representing the light cloud at rest. The IW function writes then simply (e.g. in the special case $j = 0$) : 

\beq
\label{1bise}
\xi(v.v') =\ <U(B_{v'})\phi_0|U(B_v)\phi_0>
\eeq
where $B_v$ and $B_{v'}$ are the corresponding boosts. Notice that we are dealing precisely with the Lorentz group and not with the usual Poincar\' e group. This is due to the fact that we are working within  the heavy quark limit of QCD.\par
Another important aspect worth to underline is that the light cloud belongs to a Hilbert space and that therefore the corresponding sum rules are {\it convergent}, although this feature does not survive the inclusion of radiative corrections involving {\it hard gluons}.\par
Let us now go back to previous work on the sum rule method. In the meson case $\overline{B} \to D^{(*)}\ell \overline{\nu}_{\ell}$, in the leading order of the heavy quark expansion, Bjorken sum rule (SR) \cite{4r}\cite{5r} gives the lower bound for the derivative of the meson elastic IW function at zero recoil $\rho^2 = - \xi ' (1) \geq {1\over 4}$. A new SR was formulated by Uraltsev in the heavy quark limit \cite{6r} that, combined with Bjorken's, gave the much stronger lower bound $\rho^2 \geq {3 \over 4}$. A basic ingredient in deriving this bound was the consideration of the non-forward amplitude $ \overline{B}(v_i) \to D^{(n)}(v') \to  \overline{B}(v_f)$, allowing for general four-velocities $v_i$, $v_f$, $v'$.\par
In \cite{7r} we did develop a manifestly covariant formalism within the Operator Product Expansion (OPE) and the non-forward amplitude, using the whole tower of heavy meson states \cite{2r}. We did recover Uraltsev SR plus a general class of SR that allow to bound also higher derivatives of the IW function. In particular, we found a bound on the curvature in terms of the slope $\rho^2$, namely $\xi '' (1) \geq {1 \over 5} \left [ 4 \rho^2 + 3(\rho^2)^2 \right ]$.\par
Recently, we have extended the sum rule method to the baryon IW function $\xi_{\Lambda}(w)$ of the transition $\Lambda_b \to \Lambda_c \ell \overline{\nu}_{\ell}$ \cite{8r}. We have recovered the lower bound for the slope $\rho_\Lambda^2 = - \xi '_\Lambda (1) \geq 0$ \cite{9r}, and we have generalized it by demonstrating $(-1)^n\xi_\Lambda^{(n)}(1) \geq 0$. Moreover, exploiting systematically the sum rules, we got an improved lower bound for the curvature in terms of the slope, 

\beq
\label{2e}
\sigma_\Lambda^2 = \xi ''_\Lambda (1) \geq {3 \over 5} [\rho_\Lambda^2 + (\rho_\Lambda^2)^2]
\eeq
This bound can be useful to constrain the shape of the differential spectrum of future precise data, hopefully at LHCb, on the baryon semileptonic decay $\Lambda_b \to \Lambda_c \ell \overline{\nu}_{\ell}$, that has a large measured branching ratio of about 5 \%.\par
To simplify the notation of the present paper, that is restricted to the baryon $j = 0$ case, we replace from now on for the IW function $\xi_\Lambda(w)$ of \cite{8r} by $\xi(w)$. However for the slope $\rho_\Lambda^2$ and the curvature $\sigma_\Lambda^2$ we still keep this notation, that is used below only in Sections 6 and 10. Indeed, there there could be an ambiguity in what follows below ($\rho$ labels also the irreducible representations of the Lorentz group).\par
The much more powerful method of the present paper will provide a new insight on the physics of QCD in the heavy quark limit and on its Lorentz group structure.\par
We will see that we recover the bound (\ref{2e}) and this systematic method will allow us to find bounds for higher derivatives. We will also demonstrate that if for example the bound (\ref{2e}) is saturated, then the IW function $\xi(w)$ is completely determined and given by an explicit elementary function, dependent on a single parameter. There is a simple group theoretical argument that explains this feature.\par
We restrict here to the case $j = 0$ that, interestingly, turns out to be more involved than the meson case $j = {1\over2}$ from the point of view of the decomposition of the corresponding unitary representation of the Lorentz group into irreducible ones. The study of the $j = {1\over2}$ case, that is more complicated from the spin point of view, is postponed to future work.

\section{The Lorentz group and the heavy quark limit of QCD} \hspace*{\parindent}
In the heavy mass limit, the states of a heavy hadron $H$ containing a heavy quark $Q$ is described as follows \cite{2r}

\beq
\label{3e}
|H(v),\mu,M>\ = |Q(v),\mu> \otimes\ |v,j,M>
\eeq
where there is factorization into the heavy quark state factor $|Q(v),\mu>$ and a light cloud component $|v,j,M>$ (also called "brown muck"). The velocity $v$ of the heavy hadron H is the same as the velocity of the heavy quark $Q$, and is unquantized (this is the superselection rule of \cite{9bisr}). The heavy quark $Q$ state depends only on a spin $\mu = \pm {1\over2}$ quantum number, and so belongs to a 2-dimensional Hilbert space. The light component is the complicated thing, but it does not depend on the spin state $\mu$ of the heavy quark $Q$, nor on its mass, and this gives rise to the symmetries of the heavy quark theory.\par
As advanced in the Introduction, the matrix element of a heavy-heavy current $J$ (acting only on the heavy quark) writes then

$$<H'(v'),\mu',M'|J|H(v),\mu,M>\ =\ <Q'(v'),\mu'|J|Q(v),\mu>$$
\beq
\label{4e}
<v',j',M'|v,j,M> 
\eeq
and the IW functions are defined as the coefficients, depending only on $v.v'$, in the expansion of the unknown scalar products $<v',j',M'|v,j,M>$ into independent scalars constructed from $v$, $v'$ and the polarization tensors describing the spin states of the light components.\par
Now, the crucial point in the present work is that {\it the states of the light components make up a Hilbert space in which acts a unitary representation of the Lorentz group}. In fact, this is more or less implicitly stated, and used in the literature \cite{2r}.\par

\subsection{Physical picture of a heavy quark} \hspace*{\parindent}

To see the point more clearly, let us go into the physical picture which is at the basis of (\ref{3e}). Considering first a heavy hadron {\it at rest}, with velocity

\beq
\label{4bise}
v_0 = (1,0,0,0)
\eeq
its light component is submitted to the interactions between the light particles, light quarks, light antiquarks and gluons, and to the external chromo-electric field generated by the heavy quarks at rest. This chromo-electric field does not depend on the spin $\mu$ of the heavy quark nor on its mass. We shall then have a complete orthonormal system of energy eigenstates $|v_0,j,M,\alpha>$ of the light component, where $j$ and $M$ are the angular momentum quantum numbers, and $\alpha$ is any needed additional quantum number,  

\beq
\label{5e}
<v_0,j',M',\alpha'|v_0,j,M,\alpha>\ = \delta_{j,j'} \delta_{M,M'} \delta_{\alpha,\alpha'}   
\eeq
Now, for a heavy hadron moving with a velocity $v$, the only thing which changes for the light component is that the external chromo-electric field generated by the heavy quark at rest is replaced by the external chromo-electromagnetic field generated by the heavy quark moving with the velocity $v$. Neither the Hilbert space describing the possible states of the light component, nor the interactions between the light particles, are changed. We shall then have {\it a new complete orthonormal system} of energy eigenstates $|v,j,M,\alpha>$, in the same Hilbert space. Then, because the colour fields generated by a heavy quark for different velocities are related by Lorentz transformations, we may expect that the energy eigenstates of the light component will, for various velocities, be themselves related by Lorentz transformations acting in their Hilbert space.\par

\subsection{Lorentz representation from covariant overlaps} \hspace*{\parindent}

Let us now show that such a representation of the Lorentz group does in fact underly the work of ref. \cite{2r}.\par
For integer spin $j$, the spin state of the light component is described by a polarization tensor $\epsilon^{\mu_1,...,\mu_j}$ subject to the constraints of symmetry, transversality and tracelessness

\beq
\label{6e}
v_{\mu_1} \epsilon^{\mu_1,...\mu_j}  = 0     \qquad\qquad\qquad	 g_{\mu\nu}\ \epsilon^{\mu,\nu,\mu_3,...,\mu_j} = 0
\eeq
For half-integer spin $j$, the polarization tensor becomes a Rarita-Schwinger tensor-spinor $\epsilon^{\mu_1,...\mu_{j-1/2}}_\alpha$ subject to the constraints of symmetry, transversality and tracelessness as above, and

\beq
\label{7e}
({/\hskip - 2 truemm v})_{\alpha\beta} \epsilon^{\mu_1,...,\mu_{j-1/2}}_\beta  = 0    \qquad\qquad	 (\gamma_{\mu_1})_{\alpha\beta} \epsilon^{\mu_1,...,\mu_{j-1/2}}_\beta = 0
\eeq
Then a scalar product $<v',j',\epsilon',\alpha'|v,j,\epsilon,\alpha>$ is a covariant function of the vectors $v$ and $v'$ and of the tensors (or tensor-spinors) $\epsilon'^*$ and $\epsilon$, bilinear with respect to $\epsilon'^*$ and $\epsilon$, and the IW functions, functions of the scalar $v.v'$, are introduced accordingly.\par
Now, the covariance property of the scalar products is explicitly expressed by the equality

\beq
\label{8e}
<\Lambda v',j',\Lambda \epsilon',\alpha'|\Lambda v,j,\Lambda \epsilon,\alpha>\ =\ <v',j',\epsilon',\alpha'|v,j,\epsilon,\alpha> 
\eeq 
valid for any Lorentz transformation $\Lambda$, with the transformation of a tensor (or tensor-spinor) $\epsilon$ given by

\beq
\label{9e}
(\Lambda \epsilon)^{\mu_1,...,\mu_j} = \Lambda^{\mu_1}_{\nu_1} ...  \Lambda^{\mu_j}_{\nu_j}\ \epsilon^{\nu_1,...,\nu_j}\qquad\qquad\qquad
\eeq 
\beq
\label{10e}
(\Lambda \epsilon)^{\mu_1,...,\mu_{j-1/2}}_\alpha = \Lambda^{\mu_1}_{\nu_1} ...  \Lambda^{\mu_{j-1/2}}_{\nu_{j-1/2}} D(\Lambda)_{\alpha\beta}\ \epsilon^{\nu_1,...,\nu_{j-1/2}}_\beta
\eeq

Then, let us {\it define} the operator $U(\Lambda)$, in the space of the light component states, by

\beq
\label{11e}
U(\Lambda)|v_0,j,\epsilon,\alpha>\ = |\Lambda v_0,j,\Lambda\epsilon,\alpha> 
\eeq 
where here $v_0$ is a fixed, arbitrarily chosen velocity. Let us show that $U(\Lambda)$ gives a unitary representation of the Lorentz group.\par 
Eq. (\ref{8e}) implies that $U(\Lambda)$ is a {\it unitary operator}. Let us find the action of $U(\Lambda)$ on the state $|v,j,\epsilon,\alpha>$. Using a complete orthonormal set $|v_0,j,\epsilon^{(M)},\alpha>$, where $\epsilon^{(M)}$ for $-j \leq M \leq j$ is a basis for the polarization tensors, we have
$$U(\Lambda)|v,j,\epsilon,\alpha>$$
$$= \sum_{j',M,\beta}<v_0,j',\epsilon^{(M)},\beta|v,j,\epsilon,\alpha> U(\Lambda) |v_0,j',\epsilon^{(M)},\beta>$$
$$= \sum_{j',M,\beta}<v_0,j',\epsilon^{(M)},\beta|v,j,\epsilon,\alpha> |\Lambda v_0,j',\Lambda\epsilon^{(M)},\beta>$$

Using (\ref{8e}), one gets
$$U(\Lambda)|v,j,\epsilon,\alpha>$$
$$= \sum_{j',M,\beta}<\Lambda v_0,j',\Lambda \epsilon^{(M)},\beta |\Lambda v,j, \Lambda \epsilon,\alpha> |\Lambda v_0,j', \Lambda \epsilon^{(M)},\beta> $$
and using the fact that the set $|\Lambda v_0,j, \Lambda \epsilon^{(M)},\alpha>$ is orthonormal (see (\ref{8e}) again), and complete as well, we finally obtain

\beq
\label{12e}
U(\Lambda)|v,j,\epsilon,\alpha>\ = |\Lambda v,j,\Lambda \epsilon,\alpha> 
\eeq 
From (\ref{12e}), it is easy to see that the group property $U(\Lambda ')U(\Lambda) = U(\Lambda '.\Lambda)$ holds. Therefore, we have indeed a unitary representation of the Lorentz group in the Hilbert space of the light component states.\par

\subsection{From a Lorentz representation to Isgur-Wise functions} \hspace*{\parindent}

We have shown above how a unitary representation of the Lorentz group emerges from the usual treatment of heavy hadrons in the heavy quark theory. For the present purpose, we need to go in the opposite way, namely, to show how, starting from a unitary representation of the Lorentz group, the usual treatment of heavy hadrons and the introduction of the IW functions emerges. What follows is not restricted to the $j = 0$ case, but concerns any IW function.\par
So, let us consider some unitary representation $\Lambda \to U(\Lambda)$ of the Lorentz group, or more precisely of the group $SL(2,C)$, in a Hilbert space $\mathcal{H}$. We have to identify states in $\mathcal{H}$, depending on a velocity $v$, and which are transformed according to (\ref{12e}). The difficulty is that, in the present abstract setting, we apparently have nothing like a velocity $v$ in sight. But we have in $\mathcal{H}$ an additional structure, namely the energy operator of the light component {\it for a heavy quark at rest}. Since this energy operator is invariant under rotations, we have to consider the subgroup $SU(2)$ of $SL(2,C)$. By restriction, the representation in $\mathcal{H}$ of $SL(2,C)$ gives a representation $R \to U(R)$ of $SU(2)$, and the decomposition of $\mathcal{H}$ into irreducible representations of $SU(2)$ comes into play. We then have the eigenstates $|v_0,j,M,\alpha>$ of the energy operator, classified by the angular momentum number $j$ of the irreducible representations of $SU(2)$, and {\it associated with the rest velocity} $v_0$, since their physical meaning is to describe the energy eigenstates of the light component for a heavy quark at rest.\par
Let us pursue our task, which is to express the states $|v,j,\epsilon,\alpha>$ in terms of the states $|v_0,j,M,\alpha>$. We begin with $v = v_0$. For fixed $j$ and $\alpha$, the states $|v_0,j,M,\alpha>$ constitute, for $-j \leq M \leq j$, a standard basis of a representation $j$ of $SU(2)$ :  

\beq
\label{13e}
U(R)\ |v_0,j,M,\alpha>\ = \sum_{M'}\ D^j_{M',M}(R)\ |v_0,j,M',\alpha> 
\eeq
where the rotation matrix elements $D^j_{M',M}$ are defined by

\beq
\label{40e}
D^j_{M',M} =\ <j,M'|U_j(R)|j,M>   \qquad\qquad  R \in SU(2)
\eeq

On the other hand, the states $|v_0,j,\epsilon,\alpha>$ constitute, when $\epsilon$ goes over all polarization tensors (or tensor-spinors), the whole space of a representation $j$ of $SU(2)$. In fact, when $v = v_0$, the first constraint in (\ref{6e}) and (\ref{7e}) means the vanishing of any component of $\epsilon$ if some $\mu _i$ is 0 (or if $\alpha$ = 3 or 4, in the case of a tensor-spinor), so that $\epsilon$ can be considered as a tensor (resp. tensor-spinor) constructed from ordinary 3-dimensional space (resp. and from two-dimensional spinor space). The group $SU(2)$ acts on these tensors (or tensor-spinors) in the usual way, through rotations in 3-dimensional space and through the spin $1/2$ representation in the spinor space. This representation of $SU(2)$ in the space of 3-tensors (or 3-tensor-spinors) is not irreducible, but contains an irreducible subspace of spin $j$, which is precisely the polarization 3-tensor (or 3-tensor-spinor) space selected by the other constraints, symmetry and second constraint in (\ref{6e}) and (\ref{7e}).\par
The conclusion of this rather long description is that we may introduce a standard basis $\epsilon^{(M)}$, $-j \leq M \leq j$, for the $SU(2)$ representation of spin $j$ in the space of polarization 3-tensors (or 3-tensor-spinors). This basis will be used here to demonstrate the equation (\ref{19e}) below.Then, using the notation (\ref{9e}), (\ref{10e}) for $R\epsilon$, one has

\beq
\label{14e}
R\epsilon^{(M)} = \sum_{M'} D^j_{M',M}(R)\ \epsilon^{(M')} 
\eeq
and since, according to (\ref{12e}), we want $U(R)|v_0,j,\epsilon,\alpha>\ = |v_0,j,R\epsilon,\alpha>$, the states $|v_0,j,\epsilon^{(M)},\alpha>$ are readily identified :

\beq
\label{15e}
|v_0,j,\epsilon^{(M)},\alpha>\ = |v_0,j,M,\alpha> 
\eeq
Notice that this is an identity between states described by polarization tensors $\epsilon^{(M)}$ and states belonging to a basis standard under rotations.\par
Next, denoting by $\epsilon_M$ the standard components of an arbitrary $\epsilon$, which are just the components with respect to the basis $\epsilon^{(M)}$ :

\beq
\label{16e}
\epsilon = \sum_{M} \epsilon_M \epsilon^{(M)} 
\eeq
and, from  (\ref{15e}) and  (\ref{16e}), the state $|v_0,j,\epsilon,\alpha>$ for an arbitrary $\epsilon$ is obtained  

\beq
\label{17e}
|v_0,j,\epsilon,\alpha>\ = \sum_{M} \epsilon_M\ |v_0,j,M,\alpha>
\eeq

Finally, we have to find the states $|v,j,\epsilon,\alpha>$ associated to an arbitrary velocity $v$. Wanting
(\ref{12e}) to be satisfied, we have no choice. Indeed, (\ref{12e}) gives

\beq
\label{18e}
U(\Lambda)|v_0,j,\epsilon,\alpha>\ = |\Lambda v_0,j,\Lambda \epsilon,\alpha> 
\eeq
for any Lorentz transformation $\Lambda$, so that we must have

\beq
\label{19e}
|v,j,\epsilon,\alpha>\ = \sum_{M} (\Lambda^{-1}\epsilon)_M\ U(\Lambda) |v_0,j,M,\alpha>
\eeq
for $\Lambda$ such that $\Lambda.v_0 = v$, with $v_0$ given by (\ref{4bise}).
 
Equation (\ref{19e}) is {\it our final result} here, defining, in the Hilbert space $\mathcal{H}$ of a unitary representation of $SL(2,C)$, the states $|v,j,\epsilon,\alpha>$ whose scalar products define the IW functions, in terms of $|v_0,j,M,\alpha>$ which occur as $SU(2)$ multiplets in the restriction to $SU(2)$ of the $SL(2,C)$ representation.
 
 However, in order that (\ref{19e}) be really a definition of $|v,j,\epsilon,\alpha>$, there is still something to be verified, namely that $|v,j,\epsilon,\alpha>$ does not depend on the choice of the Lorentz transformation $\Lambda$ such that $\Lambda.v_0 = v$. So, let $\Lambda '$ be another Lorentz transformation such that $\Lambda '.v_0 = v$. Since $\Lambda^{-1}\Lambda '.v_0 = v_0$,  $\Lambda^{-1}\Lambda '$ is a rotation $R \in SU(2)$, and we have $\Lambda ' = \Lambda R$. Then,
 
 $$\sum_{M} (\Lambda '^{-1}\epsilon)_M\ U(\Lambda ') |v_0,j,M,\alpha>\ = \sum_{M} (R^{-1}\Lambda^{-1}\epsilon)_M\ U(\Lambda R) |v_0,j,M,\alpha>$$
$$= \sum_{M} (R^{-1}(\Lambda^{-1}\epsilon))_M\ U(\Lambda)U(R) |v_0,j,M,\alpha>$$
Using (\ref{13e}) and (\ref{14e}) to expand $U(R) |v_0,j,M,\alpha>$ and $(R^{-1}(\Lambda^{-1}\epsilon))_M$, one obtains

 $$\sum_{M} (\Lambda '^{-1}\epsilon)_M\ U(\Lambda ') |v_0,j,M,\alpha>$$ 
 $$= \sum_{M,M',M''}D^j_{M,M'}(R^{-1})\ (\Lambda^{-1}\epsilon)_{M'}D^j_{M'',M}(R)\ U(\Lambda) |v_0,j,M'',\alpha>$$
 The sum over M is done using the group property of the $D^j_{M,M'}$ :
 
 $$\sum_{M}D^j_{M,M'}(R^{-1})D^j_{M'',M}(R) = \delta _{M'',M'}$$
 and one obtains
 
  $$\sum_{M} (\Lambda '^{-1}\epsilon)_M U(\Lambda ') |v_0,j,M,\alpha>\ = \sum_{M'} (\Lambda^{-1}\epsilon)_{M'} U(\Lambda) |v_0,j,M',\alpha>\ = |v,j,\epsilon,\alpha>$$
  proving that the state $|v,j,\epsilon,\alpha>$ defined by (\ref{19e}) does not depend on the choice of $\Lambda$ (as long as $\Lambda.v_0 = v$).
 
 To be complete, we have still to show that eq.(\ref{8e}) and (\ref{12e}) hold for these states $|v,j,\epsilon,\alpha>$. The proof of (\ref{12e}) is straightforward. Let $\Lambda$ be any Lorentz transformation. From the definition (\ref{19e}), one has
 
 $$U(\Lambda)|v,j,\epsilon,\alpha>\ = \sum_{M}  (\Lambda '^{-1}\epsilon)_{M}\ U(\Lambda)U(\Lambda ') |v_0,j,M,\alpha>$$
 $$=  \sum_{M} ((\Lambda \Lambda ')^{-1}\Lambda \epsilon)_M\ U(\Lambda \Lambda ') |v_0,j,M,\alpha>$$
 where $\Lambda '$ is some Lorentz transformation such that $\Lambda '.v_0 = v$. Then, since the Lorentz transformation $\Lambda \Lambda '$ satisfies $\Lambda \Lambda '.v_0 = \Lambda v$, using again the definition (\ref{19e}), one has
 
\beq
\label{20e}
U(\Lambda)|v,j,\epsilon,\alpha>\ = |\Lambda v,j,\Lambda \epsilon,\alpha>
\eeq
Finally, eq.(\ref{8e}), which is crucial for the definition of the IW functions, that writes here

\beq
\label{21e}
<\Lambda v',j',\Lambda \epsilon',\alpha'|\Lambda v,j,\Lambda \epsilon,\alpha>\ =\ <v',j',\epsilon',\alpha'|v,j,\epsilon,\alpha> 
\eeq
is an immediate consequence of (\ref{12e}) and of the unitarity of $U(\Lambda)$, here assumed from the start.

\section{Decomposition into irreducible representations and integral formula for the Isgur-Wise function (in the case $j = 0$)} \hspace*{\parindent}

Let us first explain, in general terms, the decomposition of unitary representations into irreducible ones, and how this gives a general integral formula for the IW functions. As it is well known, in the case of a compact group (as $SU(2)$), any unitary representation can be written as a direct sum of irreducible ones. In the present case of $SL(2,C)$ (a non-compact group), the more general notion of a direct integral is required \cite{10r}.\par
Let us denote by $X$ the set of (equivalence classes of) irreducible unitary representations of $SL(2,C)$, by $\mathcal{H}_\chi$ the Hilbert space of a representation $\chi \in X$, and by $U_\chi(\Lambda)$ the unitary operator acting in $\mathcal{H}_\chi$ which corresponds to any $\Lambda \in SL(2,C)$. Then, for any unitary representation of $SL(2,C)$, the Hilbert space $\mathcal{H}$ can be written in the form

\beq
\label{22e}
\mathcal{H} = \int_{X}^{\oplus} \oplus_{n_\chi} \mathcal{H}_\chi\ d\mu(\chi)
\eeq
where $\oplus$ on the integral sign indicates a direct integral of Hilbert spaces, $\mu$ is an arbitrary {\it positive} measure on the set $X$ and $n_\chi$ is a function on $X$ with $\geq 1$ integer values or possibly $\infty$. This is a rather symbolic formula. Explicitly, an element $\psi \in \mathcal{H}$ is a function

\beq
\label{23e}
\psi : \chi \in X \to \psi_\chi = (\psi_{1,\chi},... ,\psi_{n_\chi,\chi})  \in \oplus_{n_\chi} \mathcal{H}_\chi
\eeq
which assigns to each $\chi \in X$ an element $\psi_\chi \in \oplus_{n_\chi} \mathcal{H}_\chi$, and which is $\mu$-measurable and square $\mu$-integrable. The scalar product in $\mathcal{H}$ is given by :

\beq
\label{24e}
<\psi '|\psi>\ = \int_{X} <\psi_\chi '|\psi_\chi> d\mu(\chi)
\eeq
and the operator $U(\Lambda)$ of the representation in the space $\mathcal{H}$ is given by :

\beq
\label{25e}
\left (U(\Lambda\right )\psi)_{k,\chi} = U_\chi(\Lambda)\psi_{k,\chi}
\eeq

The more familiar notion of a direct sum is the particular case of a direct integral when the measure $\mu$ is a sum of Dirac $\delta$ functions.\par
Let us see now the consequences for the IW functions. For simplicity, we take here the case of a scalar ($j = 0$) light component. For the hadron at rest, the light component will be described by {\it some} element $\psi_0 \in \mathcal{H}$ {\it which is scalar} for the subgroup $SU(2)$ of $SL(2,C)$. Then, according to the law of transformation (\ref{25e}), requiring that $\psi_0$ is a scalar under rotations is the same as requiring that $\psi_{0,k,\chi}$ is a scalar under rotations for all $\chi$'s (technically, for $\mu$-almost all $\chi$'s) and all $k = 1,... , n_\chi$ (all $k \geq 1$ if $n_\chi = \infty$, we omit hereafter to specify this case). We therefore have to look at the $SU(2)$ scalars in each irreducible representation of $SL(2,C)$. More generally, the decomposition of the irreducible representations of $SL(2,C)$ into irreducible representations of $SU(2)$ is known (see Section below). The decomposition is by a direct sum (since $SU(2)$ is compact), and therefore each $\mathcal{H}_\chi$ admits an orthonormal basis adapted to $SU(2)$. Moreover, it turns out that each representation $j$ of $SU(2)$ appears with multiplicity 0 or 1. Then, there is a subset $X_0 \subset X$ of irreducible representations of $SL(2,C)$ containing a non-zero $SU(2)$ scalar subspace and, for $\chi \in X_0$, there is a unique (up to a phase) normalized $SU(2)$ scalar element in $\mathcal{H}_\chi$, which we denote $\phi_{0,\chi}$. Each scalar element in $\mathcal{H}_\chi$ is then proportional to $\phi_{0,\chi}$. So, one has

\beq
\label{26e}
\psi_{0,\chi} = (c_{1,\chi}\ \phi_{0,\chi},... ,c_{n_\chi,\chi}\ \phi_{0,\chi})
\eeq
with some coefficients $c_{1,\chi},... ,c_{n_\chi,\chi}$.
From the scalar product  (\ref{24e}) in $\mathcal{H}$, one sees that the normalization $<\psi_0|\psi_0>\ = 1$ of the light component amounts to

\beq
\label{27e}
\int_{X_0}\ \sum_{k=1}^{n_\chi}\ |c_{k,\chi}|^2\ d\mu(\chi) = 1
\eeq

Now, particularizing (\ref{19e}) for $j = 0$, the IW function $\xi(w)$ is given by :

\beq
\label{26bise}
\xi(w) =\ <\psi_0|U(\Lambda)\psi_0>
\eeq
where here (case $j = 0$) $\Lambda \in SL(2,C)$ is any transformation converting the rest velocity $v_0$ into a velocity $v$ with $v^0 = w$, for instance the boost along $Oz$ 

$$\Lambda_\tau = \left( \begin{array}{cc} e^{\tau/2} & 0 \\ 0 & e^{-\tau/2} \end{array} \right)	\qquad\qquad\qquad	w = ch(\tau)$$

From (\ref{26bise}), using formula (\ref{25e}) for $U(\Lambda)$ and (\ref{24e}) for the scalar product in $\mathcal{H}$, one readily obtains

\beq
\label{28e}
\xi(w) = \int_{X_0}\ \sum_{k=1}^{n_\chi}\ |c_{k,\chi}|^2\ <\phi_{0,\chi}|U_\chi(\Lambda)\phi_{0,\chi}> d\mu(\chi)
\eeq
It is then useful to introduce a notation  

\beq
\label{29e}
\xi_\chi(w) =\ <\phi_{0,\chi}|U_\chi(\Lambda)\phi_{0,\chi}>
\eeq
so that $\xi_\chi(w)$ may be called the {\it irreducible Isgur-Wise function} corresponding to $\chi$.
Introducing also the measure  

\beq
\label{30e}
d\nu(\chi) =\ \sum_{k=1}^{n_\chi}\ |c_{k,\chi}|^2\ d\mu(\chi)
\eeq
formula (\ref{28e}) writes 

\beq
\label{31e}
\xi(w) = \int_{X_0} \xi_\chi(w)\ d\nu(\chi)
\eeq
and exhibits the IW function as a mean value of the irreducible IW functions, with respect to some {\it positive} normalized measure $\nu$ :

\beq
\label{32e}
\int_{X_0} d\nu(\chi) = 1
\eeq

As we will see below, the irreducible IW function $\xi_\chi(w)$, which is the special case of  (\ref{31e}) when $\nu$ is a $\delta$ function, could be interesting as the limiting case of a heavy hadron with light component in the irreducible representation $\chi \in X$. In fact, as will be seen, an example of such a limiting case is obtained when, for a given slope of $\xi(w)$, the lower bound of its curvature is saturated.

\section{Irreducible unitary representations of the Lorentz group and their decomposition under rotations} 
\subsection{An explicit form of the irreducible representations of the Lorentz group} \hspace*{\parindent}

Let us now describe an explicit form of the irreducible unitary representations of $SL(2,C)$. Their set $X$ is divided into three sets, the set $X_p$ of representations of the principal series, the set $X_s$ of representations of the supplementary series, and the one-element set $X_t$ made up of the trivial representation.\\

\underline{Principal series}. A representation $\chi = (p,n,\rho)$ in the principal series is labelled by an integer $n \in Z$ and a real number $\rho \in R$. Actually, the representations $(p,n,\rho)$ and $(p,-n,-\rho)$ (as given below) turn out to be equivalent so that, in order to have each representation only once, $n$ and $\rho$ will be restricted as follows :

$$n = 0 \qquad\qquad\qquad \rho \geq 0$$
\beq
\label{33e}
n > 0 \qquad\qquad\qquad \rho \in R 
\eeq

The Hilbert space $\mathcal{H}_{p,n,\rho}$ is made up of functions of a complex variable $z$ with the standard scalar product

\beq
\label{34e}
<\phi'|\phi>\ = \int \overline{\phi'(z)}\ \phi(z)\ d^2z
\eeq
with the measure $d^2z$ in the complex plane being simply $d^2z = d(Re z)d(Im z)$. So $\mathcal{H}_{p,n,\rho} = L^2(C,d^2z)$.\par
The unitary operator $U_{p,n,\rho}(\Lambda)$ is given by :

\beq
\label{35e}
\left(U_{p,n,\rho}(\Lambda)\phi \right)\!(z) =  \left({{\alpha-\gamma z} \over {|\alpha-\gamma z|}}\right)^n |\alpha-\gamma z|^{2i\rho-2}\ \phi\!\left({\delta z-\beta} \over {\alpha-\gamma z}\right)
\eeq
where $\alpha$, $\beta$, $\gamma$, $\delta$ are complex matrix elements of $\Lambda \in SL(2,C)$ :

\beq
\label{36e}
\Lambda = \left( \begin{array}{cc} \alpha & \beta \\ \gamma & \delta \end{array} \right) \qquad\qquad\qquad \alpha \delta - \beta \gamma = 1
\eeq

\underline{Supplementary series}. A representation $\chi = (s,\rho)$ in the supplementary series is labelled by a real number $\rho \in R$ satisfying 

\beq
\label{37e}
0 < \rho < 1
\eeq

The Hilbert space $\mathcal{H}_{s,\rho}$ is made up of functions of a complex variable $z$ with the non-standard scalar product

\beq
\label{38e}
<\phi'|\phi>\ = \int \overline{\phi'(z_1)}\ |z_1-z_2|^{2\rho-2}\ \phi(z_2)\ d^2z_1 d^2z_2
\eeq
The positivity of this scalar product (when $0 < \rho < 1$) can be seen by Fourier transforming. The Hilbert space can be obtained by completing the pre-hilbert space of continuous functions vanishing outside a bounded region.\par
The unitary operator $U_{s,\rho}(\Lambda)$ is given by :

\beq
\label{39e}
\left (U_{s,\rho}(\Lambda)\phi \right )\!(z) =  |\alpha-\gamma z|^{-2\rho-2}\ \phi\! \left( {\delta z-\beta} \over {\alpha-\gamma z}\right )
\eeq

\underline{Trivial representation}. The trivial representation $\chi = t$ is of course the one-dimensional representation, with Hilbert space $\mathcal{H}_t = C$, scalar product $<\phi'|\phi> = \overline{\phi'(z)}\phi(z)$ and unitary operator $U_t(\Lambda) = 1$.\par

The formulae above allow, with some calculations, to see that they define unitary representations of $SL(2,C)$. For a proof that these representations are irreducible and that they exhaust the unitary irreducible representations, see (Na\"{\i}mark \cite{11r}).\par

\subsection{Decomposition under the rotation group} \hspace*{\parindent}
Next we need the decomposition of the restriction to the subgroup $SU(2)$ of each irreducible unitary representation of $SL(2,C)$.\par
The decomposition is by a direct sum (a direct integral is not needed) since $SU(2)$ is compact, so that, for each representation $\chi \in X$ we have an {\it orthonormal basis} $\phi^\chi_{j,M}$ of $\mathcal{H}_\chi$ adapted to $SU(2)$. Here we denote by $j$ the spin of an irreducible representation of $SU(2)$ (having in mind the usual notation for the spin of the light component of a heavy hadron). It turns out \cite{11r} that each representation $j$ of $SU(2)$ appears in $\chi$ with multiplicity 0 or 1, so that $\phi^\chi_{j,M}$ needs no more indices, and that the values taken by $j$ are part of the integer and half-integer numbers. For $j$ fixed, the functions $\phi^\chi_{j,M}$, $-j \leq M \leq j$ are choosen as a standard basis of the representation $j$ of $SU(2)$. This leaves arbitrary the choice of a phase for each $j$. The choice in the formulae below is for simplicity. The normalization constants are computed in Appendix B.\par
It turns out \cite{11r} that the functions $\phi^\chi_{j,M}(z)$ are expressed in terms of the rotation matrix elements $D^j_{M',M}$ defined by (\ref{40e}). A matrix $R \in SU(2)$ being of the form

\beq
\label{41e}
R = \left( \begin{array}{cc} a & b \\ -\overline{b} & \overline{a} \end{array} \right) \qquad\qquad\qquad |a|^2+|b|^2 = 1
\eeq
we shall also consider $D^j_{M',M}$ as a function of $a$ and $b$ (satisfying $|a|^2+|b|^2 = 1$). Then one has the following simple generating function 

$$\sum_{M,M'} {D^j_{M',M}(a,b) \over \sqrt{(j-M)!(j+M)!(j-M')!(j+M')!}} s^{j+M} s'^{j+M'}$$
\beq
\label{42e}
= {1 \over (2j)!} \left(bs'+\overline{a}+(as'-\overline{b})s\right)^{2j}
\eeq
and the following explicit formula :

$$D^j_{M',M}(a,b) = \sqrt{{(j-M')!(j+M')! \over (j-M)!(j+M)!}}$$
\beq
\label{43e}
\sum_{k}\ (-1)^k  \left( \begin{array}{c} j+M \\ k \end{array} \right) \left( \begin{array}{c} j-M \\ j-M'-k \end{array} \right) a^{j+M-k}\ \overline{a}^{j-M'-k}\ b^{-M+M'+k}\  \overline{b}^k 
\eeq 
The well known scalar product in the space $L^2(SU(2),dR)$ of the rotation matrix elements will also be very useful :

\beq
\label{44e}
\int D^{J_2}_{M'_2,M_2}(R)^* D^{J_1}_{M'_1,M_1}(R)\ dR = {1 \over 2J_1+1}\ \delta_{M'_1,M'_2}\ \delta_{M_1,M_2}\ \delta_{J_1,J_2} 
\eeq
(dR is the normalized invariant measure on the group $SU(2)$).\par
We can now give explicit formulae for the orthonormal basis $\phi^\chi_{j,M}$ of $\mathcal{H}_\chi$.\par

\underline{Principal series}. The spins $j$ which appear in a representation $\chi = (p,n,\rho)$ of the principal series are\par
- All the integers $j \geq {n \over 2}$  when $n$ is even.\par
- All the half-integers $j \geq {n \over 2}$ when $n$ is odd.\par
Such a spin appears with multiplicity 1. The basis functions $\phi^{p,n,\rho}_{j,M}(z)$ are :

\beq
\label{45e}
\phi^{p,n,\rho}_{j,M}(z) =  {\sqrt{2j+1} \over \sqrt{\pi}}\ (1+|z|^2)^{i\rho-1} D^j_{n/2,M}\! \left( {1 \over \sqrt{1+|z|^2}}, - {z \over \sqrt{1+|z|^2}} \right)	
\eeq
or, using the explicit formula for $D^j_{n/2,M}$ :

$$\phi^{p,n,\rho}_{j,M}(z) = {\sqrt{2j+1} \over \sqrt{\pi}}\ (-1)^{n/2-M}\sqrt{{(j-n/2)!(j+n/2)!} \over {(j-M)!(j+M)!}}\ (1+|z|^2)^{i\rho-j-1}$$ 
\beq
\label{46e}
\sum_{k}\ (-1)^k  \left( \begin{array}{c} j+M \\ k \end{array} \right) \left( \begin{array}{c} j-M \\ j-n/2-k \end{array} \right) z^{n/2-M+k}\ \overline{z}^k 
\eeq
where the range for $k$ can be limited to $0 \leq k \leq{j-n/2}$ due to the binomial factors.\par

\underline{Supplementary series}. The spins $j$ which appear in a representation $\chi = (s,\rho)$ of the supplementary series are\par
- All the integers $j \geq 0$.\par
Such a spin appears with multiplicity 1. The basis functions $\phi^{s,\rho}_{j,M}(z)$ are :

$$\phi^{s,\rho}_{j,M}(z) = {\sqrt{2j+1} \over \pi} \sqrt{{\Gamma(j+\rho+1)\Gamma(1-\rho)} \over {\Gamma(j-\rho+1)\Gamma(\rho)}}$$ 
\beq
\label{47e}
(1+|z|^2)^{-\rho-1}D^j_{0,M}\! \left( {1 \over \sqrt{1+|z|^2}}, - {z \over \sqrt{1+|z|^2}} \right) 
\eeq
or, using the explicit formula for $D^j_{0,M}$ :

$$\phi^{s,\rho}_{j,M}(z) = {\sqrt{2j+1} \over \pi}\ (-1)^M \sqrt{{\Gamma(j+\rho+1)\Gamma(1-\rho)} \over {\Gamma(j-\rho+1)\Gamma(\rho)}}{j! \over \sqrt{(j-M)!(j+M)!}}$$ 
\beq
\label{48e}
(1+|z|^2)^{-\rho-j-1}\sum_{k}\ (-1)^k  \left( \begin{array}{c} j+M \\ k \end{array} \right) \left( \begin{array}{c} j-M \\ j-k \end{array} \right) z^{-M+k}\ \overline{z}^k 
\eeq
 
\underline{Trivial representation}. Of course the only spin $j$ appearing in the trivial representation $\chi = t$ is $j = 0$, and $\phi^t_{0,0}$ is any normed element of the one-dimensional Hilbert space $\mathcal{H}_t$.

 \section{Representations relevant to the $j = 0$ case and explicit integral formula for the Isgur-Wise function} \hspace*{\parindent}
 
 Let us return to the case $j = 0$ of a scalar light component. According to the description above, the subset $X_0$ of irreducible representations of $SL(2,C)$ containing a $SU(2)$ scalar is made of : \par
 - The subset $n = 0$ of the principal series.\par
 - All the supplementary series.\par
 - The trivial representation.\par
 
 The $j = 0$ basis element (from (\ref{46e}) and (\ref{48e})) is ($k = 0$ for $n = 0$, $j = 0$)

\beq
\label{49e}
\phi^{p,0,\rho}_{0,0}(z) = {1 \over \sqrt{\pi}} (1+|z|^2)^{i\rho-1}	\qquad\qquad \chi = (p,0,\rho) \qquad \rho \geq 0
\eeq

\beq
\label{50e}
\phi^{s,\rho}_{0,0}(z) = {\sqrt{\rho} \over \pi} (1+|z|^2)^{-\rho-1}	\qquad\qquad \chi = (s,\rho) \qquad 0 < \rho < 1
\eeq

\beq
\label{51e}
\phi^t_{0,0}(z) = 1 \qquad\qquad\qquad\qquad\qquad\qquad\qquad 	\chi = t \qquad\qquad
\eeq
The corresponding irreducible IW functions, according to (\ref{29e}), are 
 
\beq
\label{52e}
\xi_\chi(w) =\ <\phi^\chi_{0,0}|U_\chi(\Lambda_\tau)\phi^\chi_{0,0}> \qquad\qquad\qquad	w = ch(\tau)
\eeq 
with 

$$\Lambda_\tau = \left( \begin{array}{cc} e^{\tau/2} & 0 \\ 0 & e^{-\tau/2} \end{array} \right)$$ 
The transformed elements $U_\chi(\Lambda_\tau)\phi^\chi_{0,0}$ are given by (\ref{35e}) and (\ref{39e}) : 

\beq
\label{53e}
\left (U_{p,0,\rho}(\Lambda_\tau)\phi^{p,0,\rho}_{0,0} \right )\!(z) = e^{(i\rho-1)\tau}\phi^{p,0,\rho}_{0,0}(e^{-\tau}z) = {1 \over \sqrt{\pi}} (e^\tau+e^{-\tau}|z|^2)^{i\rho-1}
\eeq

\beq
\label{54e}
\left (U_{s,\rho}(\Lambda_\tau)\phi^{s,\rho}_{0,0} \right )\!(z) = e^{-(\rho+1)\tau}\phi^{s,\rho}_{0,0}(e^{-\tau}z) = {\sqrt{\rho} \over \sqrt{\pi}} (e^\tau+e^{-\tau}|z|^2)^{-\rho-1} \\
\eeq

\beq
\label{55e}
U_{t}(\Lambda_\tau)\phi^{t}_{0,0} = 1  
\eeq
and, the scalar products being given by (\ref{34e}) and (\ref{38e}), we have : 

\beq
\label{56e}
\xi_{p,0,\rho}(w) = {1 \over \pi} \int (1+|z|^2)^{-i\rho-1}(e^\tau+e^{-\tau}|z|^2)^{i\rho-1}d^2z \qquad\qquad\qquad
\eeq 

\beq
\label{57e}
\xi_{s,\rho}(w) = {\rho \over \pi^2} \int (1+|z'|^2)^{-\rho-1}\ |z'-z|^{2\rho-2}\ (e^\tau+e^{-\tau}|z|^2)^{-\rho-1}d^2z'd^2z
\eeq 

\beq
\label{58e}
\xi_{t}(w) = 1 \qquad\qquad\qquad\qquad\qquad\qquad\qquad\qquad\qquad\qquad\qquad\qquad
\eeq
The integrals being computed in Appendix C, one obtains

\beq
\label{59e}
\xi_{p,0,\rho}(w) = {sin(\rho \tau) \over {\rho\ sh(\tau)}}		\qquad\qquad\qquad     (0 \leq \rho)\qquad\qquad\qquad\qquad\qquad
\eeq 

\beq
\label{60e}
\xi_{s,\rho}(w) = {sh(\rho \tau) \over {\rho\ sh(\tau)}}		\qquad\qquad\qquad     (0 < \rho < 1)\qquad\qquad\qquad\qquad\qquad
\eeq 

\beq
\label{61e}
\xi_{t}(w) = 1 \qquad\qquad\qquad\qquad\qquad\qquad\qquad\qquad\qquad\qquad\qquad\qquad
\eeq
Then our fundamental integral formula (\ref{31e}) for the IW function writes :

\beq
\label{62e}
\xi(w) = \int_{[0,\infty[} {sin(\rho \tau) \over {\rho\ sh(\tau)}}\ d\nu_p(\rho) + \int_{]0,1[} {sh(\rho \tau) \over {\rho\ sh(\tau)}}\ d\nu_s(\rho) + \nu_t \qquad\qquad w = ch(\tau)
\eeq
where $\nu_p$ and $\nu_s$ are positive measures on [0,$\infty$[ and ]0,1[, and $\nu_t$ is a real number $\geq$ 0 (the same thing as a positive measure on the one element set \{0\}), with the only condition that 

\beq
\label{63e}
\int_{[0,\infty[} d\nu_p(\rho) + \int_{]0,1[} d\nu_s(\rho) + \nu_t = 1
\eeq
(and the precise specification of the domain of integration is needed because $\nu_p$ and $\nu_s$ may include Dirac measures).\par
For the derivatives $\xi^{(k)}(1)$, the formulae (\ref{59e})-(\ref{62e}) give

\beq
\label{64e}
\xi^{(k)}(1) = \int_{[0,\infty[} \xi^{(k)}_{p,0,\rho}(1)\ d\nu_p(\rho) + \int_{]0,1[} \xi^{(k)}_{s,\rho}(1)\ d\nu_s(\rho) + \nu_t\ \delta_{k,0} \qquad\qquad 
\eeq
with the surprisingly simple expressions for the lower derivatives obtained by direct calculation :

\beq
\label{65e}
\xi_{p,0,\rho}(1) = 1	\qquad\qquad\qquad\qquad    \xi_{s,\rho}(1) = 1 
\eeq
$$\xi'_{p,0,\rho}(1) = - {1+\rho^2 \over 3}	\qquad\qquad\qquad\qquad   \xi'_{s,\rho}(1) = - {1-\rho^2 \over 3}$$
$$\xi''_{p,0,\rho}(1) = {(1+\rho^2)(4+\rho^2) \over 15}   \qquad\qquad\qquad\qquad	  \xi''_{s,\rho}(1) =  {(1-\rho^2)(4-\rho^2) \over 15}$$
$$\xi^{(3)}_{p,0,\rho}(1) = - {(1+\rho^2)(4+\rho^2)(9+\rho^2) \over 105}	\qquad\qquad   \xi^{(3)}_{s,\rho}(1) =   - {(1-\rho^2)(4-\rho^2)(9-\rho^2) \over 105}$$
$$..........$$
and we have

\beq
\label{66e}
\xi(1) = 1	\qquad\qquad\qquad\qquad    
\eeq
$$\xi'(1) = - {1 \over 3} \left[\int_{[0,\infty[} (1+\rho^2)\ d\nu_p(\rho) + \int_{]0,1[} (1-\rho^2)\ d\nu_s(\rho)\right]$$
$$\xi''(1) = {1 \over 15} \left[\int_{[0,\infty[} (1+\rho^2)(4+\rho^2)\ d\nu_p(\rho) + \int_{]0,1[} (1-\rho^2)(4-\rho^2)\ d\nu_s(\rho)\right]$$
$$\xi^{(3)}(1) = - {1 \over 105} \left[\int_{[0,\infty[} (1+\rho^2)(4+\rho^2)(9+\rho^2)d\nu_p(\rho) + \int_{]0,1[} (1-\rho^2)(4-\rho^2)(4-\rho^2)d\nu_s(\rho)\right]$$
$$..........$$
where $\nu_p$ and $\nu_s$ are arbitrary positive measures satisfying

\beq
\label{67e}
\int_{[0,\infty[} d\nu_p(\rho) + \int_{]0,1[} d\nu_s(\rho) \leq 1
\eeq

At first sight, the deduction of the constraints on the derivatives $\xi^{(k)}(1)$ from (\ref{66e}) under the condition (\ref{67e}) promises to be a tricky work. However, the problem will be reduced to an already solved one by rewriting (\ref{62e}) in a simpler form (namely with just one integral). This is possible because the irreducible IW functions, principal, supplementary and trivial, can all be put into a {\it one parameter family} :

\beq
\label{68e}
\xi_x(w) = {sh(\tau\sqrt{1-x}) \over sh(\tau)\sqrt{1-x}} = {sin(\tau\sqrt{x-1}) \over sh(\tau)\sqrt{x-1}}
\eeq

Indeed we have

$$\xi_{p,0,\rho}(w) = \xi_{x}(w) \qquad x = 1+\rho^2, \qquad \rho \in [0,\infty[ \qquad \Leftrightarrow \qquad x \in [1,\infty[$$
$$\xi_{s,\rho}(w) = \xi_{x}(w) \qquad x = 1-\rho^2, \qquad \rho \in ]0,1[  \qquad \Leftrightarrow \qquad  x \in ]0,1[$$
\beq
\label{69e}
\xi_{t}(w) = \xi_{x}(w) \qquad\qquad\qquad\ x = 0,  \qquad\qquad\qquad\qquad x\in\{0\}
\eeq
so that $\xi_x(w)$ is $\xi_{t}(w)$ for $x = 0$, is $\xi_{s,\rho}(w)$ for $0 < x < 1$, and is $\xi_{p,0,\rho}(w)$ for $1 \leq x$. Then formula (\ref{62e}) writes :

\beq
\label{70e}
\xi(w) = \int_{[0,\infty[} \xi_x(w)\ d\nu(x)
\eeq
with $\xi_x(w)$ given explicitly by (\ref{68e}) and $\nu$ is a positive measure on $[0,\infty[$, with the only condition that 

\beq
\label{71e}
\int_{[0,\infty[} d\nu(x) = 1
\eeq
According to (\ref{69e}), $\nu$ can be obtained in terms of $\nu_t$, $\nu_s$, $\nu_p$, through its restrictions to the subsets $\{0\}$, $]0,1[$, $[1,\infty[$ of $[0,\infty[$, by changes of variable. In particular, the "trivial" $\nu_t$ term in (\ref{62e}) will be due to a Dirac $\delta$ contribution to $\nu$ at $x = 0$.\par
For the derivatives $\xi^{(k)}(1)$, the formula (\ref{70e}) gives 

\beq
\label{72e}
\xi^{(k)}(1) = \int_{[0,\infty[} \xi^{(k)}_x(1)\ d\nu(x)
\eeq
As may be suspected from (\ref{65e}), $\xi^{(k)}_x(1)$ is a polynomial of degree $k$ in $x$. As shown in Appendix D, it is given by :

\beq
\label{73e}
\xi^{(k)}_x(1) = (-1)^k\ 2^k {k! \over (2k+1)!} \prod^k_{i=1} (x+i^2-1)
\eeq

\section{The constraints on moments and on derivatives of the Isgur-Wise function} \hspace*{\parindent}
 
We may now deduce the constraints on the derivatives. From (\ref{72e}) and (\ref{73e}), the derivative 	  
$\xi^{(k)}(1)$ is given by the {\it expectation value} of a polynomial of degree $k$ :

\beq
\label{74e}
\xi^{(k)}(1) = (-1)^k\ 2^k {k! \over (2k+1)!} <\prod^k_{i=1} (x+i^2-1)>
\eeq
with the expectation value defined by

\beq
\label{75e}
< f(x) >\ = \int_{[0,\infty[} f(x)\ d\nu(x)
\eeq
for some normalized positive measure $\nu$ supported in $[0,\infty[$. One obtains

$$\xi(1) =\ <1> = 1$$
$$\xi'(1) = - {1 \over 3} <x>$$
$$\xi''(1) = {1 \over 15} <x(3+x)>$$
$$\xi^{(3)}(1) = - {1 \over 105} <x(3+x)(8+x)>$$
\beq
\label{76e}
\xi^{(4)}(1) = {1 \over 945} <x(3+x)(8+x)(15+x)>
\eeq
Expanding the polynomial, $\xi^{(k)}(1)$ is expressed as a combination of the moments $\mu_0$, $\mu_1$,... $\mu_k$ of $x$, a moment $\mu_n$ being the expectation value of $x^n$ :  

\beq
\label{76bise}
\mu_n =\  < x^n >
\eeq
(notice that these moments could be infinite for $n \geq 1$, although, in the following, we will not consider this case).\par
So we have
$$\xi(1) = \mu_0 = 1$$
$$\xi'(1) = - {1 \over 3}\ \mu_1$$
$$\xi''(1) = {1 \over 15}\ (3\mu_1+\mu_2)$$
$$\xi^{(3)}(1) = - {1 \over 105}\ (24\mu_1+11\mu_2+\mu_3)$$
\beq
\label{77e}
\xi^{(4)}(1) = {1 \over 945}\ (360\mu_1+189\mu_2+26\mu_3+\mu_4)
\eeq
These equations can be solved step by step, and the moment $\mu_k$ is expressed as a combination of the derivatives $\xi(1)$, $\xi'(1)$,... $\xi^{(k)}(1)$ :

$$\mu_0 = \xi(1) = 1$$
$$\mu_1 = - 3\ \xi'(1)$$
$$\mu_2 = 3\left[3\ \xi'(1)+5\ \xi''(1)\right]$$
$$\mu_3 = - 3\left[9\ \xi'(1)+55\ \xi''(1)+35\ \xi^{(3)}(1)\right]$$
\beq
\label{78e}
\mu_4 = 3\left[27\ \xi'(1)+485\ \xi''(1)+910\ \xi^{(3)}(1)+315\ \xi^{(4)}(1)\right]
\eeq

Now, in \cite{3r} one has obtained a whole set of constraints on the moments of a variable with positive values, and also shown that this set of constraints is optimal, meaning that it cannot be improved (there are no other nor more strict constraints from the general definition of the moments).\par
In fact, \cite{3r} was concerned only with the particular case of a measure $\nu$ of the form

$$d\nu(x) = w(x)\ dx$$
with a weight function $w$ (the measure $\nu$ is then said to be completely continuous with respect to the measure $dx$). In particular, this excludes Dirac $\delta$ contributions to $\nu$ and this, while perhaps physically reasonable, is not assumed in the present context.\par
However, the deduction of the constraints in \cite{3r} goes through in the present case of an arbitrary positive measure $\nu$ just by replacing strict inequalities $>$ by non-strict ones $\geq$. Therefore, the set of constraints is as follows. For any $n \geq 0$, one has \cite{3r}

\beq
\label{79e}
det\left[(\mu_{i+j})_{0\leq i,j \leq n}\right] \geq 0
\eeq
\beq
\label{80e}
det\left[(\mu_{i+j+1})_{0\leq i,j \leq n}\right] \geq 0
\eeq
(On the other hand, the optimality proof in \cite{3r} needs additional arguments in the present case).\par
Since each moment $\mu_k$ is a combination of the derivatives $\xi(1)$, $\xi'(1)$,... $\xi^{(k)}(1)$, the constraints on the moments translate into constraints on the derivatives.\par
We shall treat here in detail only the constraints on $\mu_1$, $\mu_2$, $\mu_3$, $\mu_4$, which are given respectively by (\ref{80e}) ($n = 0$), (\ref{79e}) ($n = 1$), (\ref{80e}) ($n = 1$), (\ref{79e}) ($n = 2$).

\beq
\label{81e}
\mu_1 \geq 0 \qquad\qquad\qquad\qquad\qquad\qquad\qquad\qquad\qquad\qquad\qquad
\eeq

\beq
\label{82e}
det \left( \begin{array}{cc}
1&\mu_1\\
\mu_1&\mu_2\\
\end{array} \right) = \mu_2 - \mu_1^2 \geq 0 \qquad\qquad\qquad\qquad\qquad\qquad
\eeq

\beq
\label{83e}
det \left( \begin{array}{cc}
\mu_1&\mu_2\\
\mu_2&\mu_3\\
\end{array} \right) = \mu_1 \mu_3 - \mu_2^2 \geq 0 \qquad\qquad\qquad\qquad\qquad\qquad
\eeq

\beq
\label{84e}
det \left( \begin{array}{ccc}
1&\mu_1&\mu_2\\
\mu_1&\mu_2&\mu_3\\
\mu_2&\mu_3&\mu_4\\
\end{array} \right) = (\mu_2 - \mu_1^2)\mu_4 - (\mu_3^2 - 2 \mu_1 \mu_2 \mu_3 + \mu_2^3) \geq 0
\eeq
Clearly, each moment $\mu_k$ is bounded from below, and the lower bound is given by (\ref{79e}) for $k$ even and by (\ref{80e}) for $k$ odd in terms of the lower moments. So (\ref{81e})-(\ref{84e}) give :

\beq
\label{85e}
\mu_1 \geq 0 \qquad\qquad\qquad\qquad\qquad\qquad\qquad\qquad\qquad
\eeq

\beq
\label{86e}
\mu_2 \geq \mu_1^2 \qquad\qquad\qquad\qquad\qquad\qquad\qquad\qquad\qquad
\eeq

\beq
\label{87e}
\mu_3 \geq {\mu_2^2 \over \mu_1} \qquad\qquad\qquad\qquad\qquad\qquad\qquad\qquad\qquad
\eeq

\beq
\label{88e}
\mu_4 \geq {\mu_3^2 - 2 \mu_1 \mu_2 \mu_3 +  \mu_2^3  \over \mu_2 - \mu_1^2} = {(\mu_3 - \mu_1 \mu_2)^2  \over \mu_2 - \mu_1^2} + \mu_2^2 \qquad
\eeq

If one of these inequalities is saturated, the inequalities following it may be meaningless. However, even in such a case, the inequalities (\ref{79e}) and (\ref{80e}) remain perfectly valid. Moreover we have in fact much more in this case, because the measure $\nu$ is then a {\it completely determined} combination of $\delta$ functions, and the IW function is completely fixed and explicitly given. In the following, we shall {\it explicitly} show this for the saturation of inequalities (\ref{85e})-(\ref{87e}).\par

So, in the case $\mu_1 = 0$ of (\ref{85e}), eq. (\ref{87e}) is meaningless. Eqs. (\ref{81e})-(\ref{84e}) are nevertheless fully valid and, for instance, (\ref{83e}) gives $\mu_2 = 0$. In fact, when $\mu_1 = 0$ we have much more, since the condition

\beq
\label{89e}
\mu_1 = \int_{[0,\infty[} x\ d\nu(x) = 0
\eeq
completely determines the measure $\nu$ :

\beq
\label{90e}
\mu_1 = 0 \qquad\qquad \Leftrightarrow \qquad\qquad d\nu(x) = \delta(x)\ dx
\eeq
(in particular then, $\mu_k = 0$ for all $k \geq 1$).\par

Also if the bound (\ref{86e}) is saturated, eq. (\ref{88e}) becomes meaningless and for instance eq. (\ref{84e}) gives $\mu_3 = \mu_1^3$. In fact, when $\mu_2 - \mu_1^2 = 0$ we have again much more, since the condition

\beq
\label{91e}
\mu_2 - \mu_1^2 = \int_{[0,\infty[} (x - \mu_1)^2\ d\nu(x) = 0
\eeq
also completely determines the measure $\nu$. Since the integrand $(x-\mu_1)^2$ is $> 0$ when $x \neq \mu_1$, the integral can vanish only is the measure is concentrated on the set $\{\mu_1\}$, and with the condition $< 1 >\ = 1$, we have :

\beq
\label{92e}
\mu_2 = \mu_1^2 \qquad\qquad \Leftrightarrow \qquad\qquad d\nu(x) = \delta(x - \mu_1)\ dx
\eeq
(in particular then, $\mu_k = \mu_1^k$ for all $k \geq 1$).\par

In the same way, if (\ref{81e}) is strict ($\mu_1 > 0$) and (\ref{83e}) is saturated ($\mu_3 = {\mu_2^2 \over \mu_1}$), we have the condition

\beq
\label{93e}
\mu_3 - {\mu_2^2 \over \mu_1} = \int_{[0,\infty[} x \left( x -  {\mu_2 \over \mu_1} \right)^2 d\nu(x) = 0
\eeq
which completely determines the measure $\nu$. The measure is concentrated on the set of two elements $\{0,{\mu_2 \over \mu_1}\}$, and with the conditions $< 1 >\ = 1$ and $< x >\ = \mu_1$, we have : 

\beq
\label{94e}
\mu_1 > 0,\  \mu_3 = {\mu_2^2 \over \mu_1} \qquad \Leftrightarrow \qquad d\nu(x) = \left [ \left (1 - {\mu_1^2 \over \mu_2} \right ) \delta(x) + {\mu_1^2 \over \mu_2}\ \delta\! \left ( x - {\mu_2 \over \mu_1}\right ) \right ] dx 
\eeq
(in particular then, $\mu_k = \left ({\mu_2 \over \mu_1} \right )^{k-1} \mu_1$ for all $k \geq 1$).\par
Incidently, this gives a hint on how to complete the optimality proof of \cite{3r}. If all the inequalities (\ref{79e}), (\ref{80e}) are strict, then the proof in \cite{3r} is valid. If there is an equality, one can show that the values of the moments can be obtained by some finite combination of Dirac $\delta$ measures for $\nu$.\par

Finally, using (\ref{78e}), the results on the moments $\mu_k$ are converted into results on the derivatives $\xi^{(k)}(1)$. With our usual notation for the slope and curvature

\beq
\label{95e}
\rho_\Lambda^2 = - \xi'(1) \qquad\qquad\qquad \sigma_\Lambda^2 = \xi''(1)
\eeq
the constraints (\ref{85e})-(\ref{88e}) write : 

\beq
\label{96e}
\rho_\Lambda^2 \geq 0 \qquad\qquad\qquad\qquad\qquad\qquad\qquad\qquad\qquad\qquad\qquad\qquad\qquad
\eeq

\beq
\label{97e}
\sigma_\Lambda^2 \geq {3 \over 5} \rho_\Lambda^2 (1 + \rho_\Lambda^2) \qquad\qquad\qquad\qquad\qquad\qquad\qquad\qquad\qquad\qquad\qquad
\eeq

\beq
\label{98e}
- \xi^{(3)}(1) \geq {5 \over 7} {\sigma_\Lambda^2 \over \rho_\Lambda^2} (\rho_\Lambda^2 + \sigma_\Lambda^2) \qquad\qquad\qquad\qquad\qquad\qquad\qquad\qquad\qquad\qquad\qquad
\eeq

\beq
\label{99e}
\xi^{(4)}(1) \geq {5 \over 63} {1 \over 5 \sigma_\Lambda^2 - 3 \rho_\Lambda^2 (1+\rho_\Lambda^2)} \qquad\qquad\qquad\qquad\qquad\qquad\qquad\qquad\qquad
\eeq 
$$ \left [3 \sigma_\Lambda^2 (\sigma_\Lambda^2 + \rho_\Lambda^2)(5 \sigma_\Lambda^2 + 8 \rho_\Lambda^2 + 8) +
7 \left ( 12 \rho_\Lambda^2 (1+\rho_\Lambda^2) + 2(3 \rho_\Lambda^2 - 2) \sigma_\Lambda^2 + 7 \xi^{(3)}(1) \right ) \xi^{(3)}(1)\right ]$$
$$ = {5 \over 63} {1 \over (\rho_\Lambda^2)^2} \sigma_\Lambda^2 (\sigma_\Lambda^2 + \rho_\Lambda^2)(5 \sigma_\Lambda^2 + 12 \rho_\Lambda^2) \qquad\qquad\qquad\qquad\qquad\qquad\qquad\qquad\qquad $$  
$$ + {5 \over 9} \left [ {2 \over \rho_\Lambda^2} (\sigma_\Lambda^2 + 2 \rho_\Lambda^2) + {7 \over 5}\left ( {- \xi^{(3)}(1)-(-\xi^{(3)}(1))_{min} \over \sigma_\Lambda^2-(\sigma_\Lambda^2)_{min}} \right ) \right ] \left [ - \xi^{(3)}(1)-(-\xi^{(3)}(1))_{min} \right ]$$\par

Eliminating $\sigma_\Lambda^2$ from (\ref{98e}), we have a looser but simpler bound for $-\xi^{(3)}(1)$ depending only on the slope :

\beq
\label{100e}
- \xi^{(3)}(1) \geq {3 \over 35}\ \rho_\Lambda^2 (1+\rho_\Lambda^2) (8 + 3\rho_\Lambda^2) 
\eeq

Eliminating  $\xi^{(3)}(1)$ from (\ref{99e}), we have also a looser but simpler bound for  $\xi^{(4)}(1)$, depending only on the slope and the curvature :

\beq
\label{101e}
\xi^{(4)}(1) \geq {5 \over 63} {1 \over (\rho_\Lambda^2)^2}\ \sigma_\Lambda^2 (\sigma_\Lambda^2 + \rho_\Lambda^2)(5 \sigma_\Lambda^2 + 12 \rho_\Lambda^2)
\eeq 
and eliminating $\sigma_\Lambda^2$ from (\ref{101e}), we have :

\beq
\label{102e}
\xi^{(4)}(1) \geq {1 \over 35} \rho_\Lambda^2 (1 + \rho_\Lambda^2)(8 + 3 \rho_\Lambda^2)(5 + \rho_\Lambda^2)
\eeq

\subsection{Illustration of the inequalities for some models of the Isgur-Wise function} \hspace*{\parindent}

To have some feeling of what happens numerically, let us illustrate the preceding inequalities for two ansatze of the IW function, the exponential and the "dipole" forms, that depend on a single parameter $c = \rho_\Lambda^2$.

In the case of the exponential form

\beq
\label{102bise}
\xi(w) = exp[-c(w-1)]
\eeq

From its derivatives 

$$(-1)^k\ \xi^{(k)}(1) = c^k$$
the inequalities (\ref{96e})-(\ref{99e}) yield respectively the following bounds on the parameter $ c = \rho_\Lambda^2 $  :

\beq
\label{103bise} 
c \geq 0, \qquad \qquad c \geq 1.5,  \qquad \qquad c \geq 2.5,  \qquad\qquad c \geq 4.28
\eeq

We observe that in this case the lower bound on the slope $c$ increases strongly as we impose the constraint coming from higher order derivatives. We can suspect therefore that this function is not physically acceptable. Indeed, we will below rigorously demonstrate that this is the case.\par
Let us now consider another ansatz, namely the "dipole" form

\beq
\label{104bise}
\xi(w) = \left ( 2 \over {w+1} \right )^{2c}
\eeq
From its derivatives 

$$(-1)^k\xi^{(k)}(1) = {2c(2c+1)...(2c+k-1) \over 2^k}$$
in this case we find, from (\ref{96e}), the trivial lower bound $c = \rho_\Lambda^2Ê\geq 0$, while from the three inequalities (\ref{97e})-(\ref{99e}) we get the same lower bound, namely

\beq
\label{105bise}
c = \rho_\Lambda^2 \geq {1 \over 4}
\eeq

Therefore it seems that, unlike the exponential form (\ref{102bise}), the "dipole" ansatz (\ref{104bise}) satisfies in a regular way the inequalities that the derivatives of the IW function must fulfill. We will demonstrate in Subsection 9.5 that this function satisfies indeed the general constraints that the IW function must fulfill for $c = \rho_\Lambda^2Ê\geq {1 \over 4}$.\par
As another example, let us consider the form

\beq
\label{105bis1e}
\xi(w) = {1 \over \left[{1+{c \over 2}(w-1)}\right]^2}
\eeq
that, as discussed below, has been proposed in the literature. The first derivatives read : 

\beq
\label{105bis2e}
\rho_\Lambda^2 = c	\qquad \qquad \qquad \sigma_\Lambda^2 = {3 \over 2} (\rho_\Lambda^2)^2 \qquad \qquad \qquad -\xi^{(3)}(1) = 3  (\rho_\Lambda^2)^3
\eeq
and the inequalities (\ref{97e})-(\ref{99e}) yield respectively the following bounds on the parameter $ c = \rho_\Lambda^2 $  :

\beq
\label{105bis3e}
\rho_\Lambda^2 \geq {2 \over 3} \qquad \qquad \qquad \rho_\Lambda^2 \geq {10 \over 13} \qquad \qquad \qquad \rho_\Lambda^2 \geq 0.86
\eeq
The successive lower bounds on the slope slowly grow and converge towards $\rho_\Lambda^2 = 1$. We will demonstrate in Subsection 9.5 that this function satisfies the general constraints that the IW function must fulfill for $c = \rho_\Lambda^2Ê\geq 1$.\par

\subsection{Completely explicit form of the Isgur-Wise function when the inequality on one of the low  order derivatives  is saturated} \hspace*{\parindent}

We have seen above that if an equality is saturated, then the measure $\nu$ is completely fixed. According to (\ref{70e}), the IW function $\xi(w)$ is also {\it completely fixed}.\par

If (\ref{96e}) is saturated, namely if $\rho_\Lambda^2 = 0$, then from (\ref{90e}) we have : 

\beq
\label{103e}
\xi(w) = 1
\eeq

This being physically excluded, we may replace (\ref{96e}) by the strict inequality $\rho_\Lambda^2 > 0$.\par

If (\ref{97e}) is saturated, namely if $\sigma_\Lambda^2 = {3 \over 5} \rho_\Lambda^2 (1 + \rho_\Lambda^2)$, then from (\ref{92e}) we have : 

\beq
\label{104e}
\xi(w) = {sh\left (\tau \sqrt{1-3\rho_\Lambda^2} \right ) \over sh(\tau) \sqrt{1 - 3\rho_\Lambda^2}} = {sin\left (\tau \sqrt{3\rho_\Lambda^2-1} \right ) \over sh(\tau) \sqrt{3\rho_\Lambda^2-1}} \qquad\qquad w = ch(\tau)
\eeq

We have here possible physically limiting cases (if $\rho_\Lambda^2 > 0$). In fact (\ref{104e}) means simply, as we have announced above, that the light component of the heavy hadron belongs to some {\it irreducible} representation of the Lorentz group.\par

In view of (\ref{69e}), when the slope $\rho_\Lambda^2$ goes from $0$ to $\infty$, each irreducible representation occurs in turn : $\rho_\Lambda^2 = 0$ for the trivial representation, $0 < \rho_\Lambda^2 < {1 \over 3}$ for the supplementary series, and $ {1 \over 3} \leq \rho_\Lambda^2$ for $n = 0$ principal series.\par 

From (\ref{73e}), the derivatives $\xi^{(k)}(1)$ are then given by :

\beq
\label{105e}
\xi^{(k)}(1) = (-1)^k\ 2^k {k! \over (2k+1)!} \prod^k_{i=1} (3 \rho_\Lambda^2+i^2-1)
\eeq

Finally, let us consider the case where (\ref{98e}) is saturated, $- \xi^{(3)} = {5 \over 7} {\sigma_\Lambda^2 \over \rho_\Lambda^2} (\rho_\Lambda^2 + \sigma_\Lambda^2)$  (and $\rho_\Lambda^2 > 0$). Then, from (\ref{94e}) we have :

\beq
\label{106e}
\xi(w) = {\mu_1^2 \over \mu_2}\ {sh\left (\tau \sqrt{1 - {\mu_2 \over \mu_1}} \right ) \over sh(\tau)  \sqrt{1 - {\mu_2 \over \mu_1}}} + \left (1 - {\mu_1^2 \over \mu_2} \right )
\eeq
where 

\beq
\label{107e}
\mu_1 = 3 \rho_\Lambda^2 \qquad\qquad\qquad \mu_2 = 3 (5 \sigma_\Lambda^2 - 3 \rho_\Lambda^2)
\eeq
and $w = ch(\tau)$. If we impose the physical condition that $\xi(w) \to 0$ when $w \to \infty$, then (\ref{106e}) is possible only if ${\mu_1^2 \over \mu_2} = 1$, that is if (\ref{97e}) is saturated, a case fully discussed just above. So, we may conclude that if the inequality (\ref{97e}) is strict, then inequality (\ref{98e}) must also be strict (to avoid a non vanishing limit of $\xi(w)$ when $w \to \infty$).   

\section{Demonstration from sum rules that the Isgur-Wise function is of positive type : application to the exponential form} \hspace*{\parindent}

\subsection{Demonstration from sum rules that the Isgur-Wise function is of positive type} \hspace*{\parindent}

In this part we will demonstrate that the IW function $\xi(w)$ is of {\it positive type}, i.e. that, for any value of $N$ and any complex numbers $a_1,..., a_N$ and velocities $v_1,..., v_N$ satisfies :

\beq
\label{108e}
\sum_{i,j=1}^N a_i^*a_j\ \xi(v_i.v_j) \geq 0
\eeq
Notice that a positive type function does not mean that this function is positive for all values of its argument.\par
As pointed out in the Introduction, in the heavy quark limit of QCD from the OPE and the non-forward amplitude, we have demonstrated a sum rule for the $j = 0$ case \cite{8r}, from which we have obtained the inequalities for the slope (\ref{96e}) \cite{9r} and for the curvature (\ref{97e}).\par
We have recently realized that the expression for this sum rule can be simplified enormously. Using the expression for the sum rule obtained in \cite{8r}, this equivalent form is deduced in Appendix E (we replace the subindex $f$ by $j$) :

$$\xi(w_{ij}) = \sum_n \sum_L \tau_L^{(n)}(w_i)^*\tau_L^{(n)}(w_j)$$
\beq
\label{109e}	
\sum_{0 \leq k \leq L/2} C_{L,k}\ (w_i^2-1)^k (w_j^2-1)^k(w_iw_j-w_{ij})^{L-2k}
\eeq 		
where $w_i = v_i.v'$, $w_j = v_j.v'$, $w_{ij} = v_i.v_j$ and $v_i$, $v_j$, $v'$ are the initial, final and intermediate state four-velocities in the sum rule, and $\tau_L^{(n)}(w)$ are the IW functions for the transition $0^+ \to L^P$ with $P=(-1)^L$, and the coefficients $C_{L,k} $ are given by

\beq
\label{110e}
C_{L,k} = (-1)^k {(L!)^2 \over (2L)!} {(2L-2k)! \over k!(L-k)!(L-2k)!}
\eeq
The last sum in (\ref{109e}) can be expressed in terms of a Legendre polynomial, as demonstrated in Appendix A of the first reference of \cite{7r}, and explicitly written in Appendix E of the present paper. We use now this derivation to express this Legendre polynomial in terms of spherical harmonics. Without loss of generality, let us use the rest frame for the intermediate states, i.e. $v' = (1,0,0,0)$, that gives

\beq
\label{111e}
w_i^2-1 = {\vec v}^{\,2}_i 	\qquad\qquad w_j^2-1 = {\vec v}^{\,2}_j \qquad\qquad w_iw_j-w_{ij} =  {\vec v}_i. {\vec v}_j 
\eeq
and using the results of Appendix A of the first reference of \cite{7r} we obtain

\beq
\label{112e}
\sum_{0 \leq k \leq L/2} C_{L,k} ({\vec v}^{\,2}_i)^k ({\vec v}^{\,2}_j)^k ({\vec v}_i. {\vec v}_j)^{L-2k} = 4\pi\ 2^L {(L!)^2 \over (2L+1)!} \sum_{m=-L}^{m=+L} \mathcal{Y}_L^m({\vec v}_i)^*\ \mathcal{Y}_L^m({\vec v}_j) 
\eeq

Combining the previous equations, we find

\beq
\label{113e}
\sum_{i,j=1}^N a_i^*a_j\ \xi(v_i.v_j) = 4\pi \sum_{i,j=1}^N \sum_n \sum_L {2^L(L!)^2 \over (2L+1)!}
\eeq
$$ \sum_{m=-L}^{m=+L} \left [ a_i\ \tau_L^{(n)}\! \left (\sqrt{1+{\vec v}^{\,2}_i} \right ) \mathcal{Y}_L^m({\vec v}_i) \right ]^* \left [ a_j\ \tau_L^{(n)}\!\left (\sqrt{1+{\vec v}^{\,2}_j} \right ) \mathcal{Y}_L^m({\vec v}_j)  \right ] \geq 0$$
and therefore the inequality (\ref{108e}) has been proved.\par
In conclusion, we have demonstrated that the IW function $\xi(w)$ is of positive type. The inequality (\ref{108e}), concerning a Riemann sum, would read, in a continuous (and covariant) form,

\beq
\label{114e}
\int {d^3{\vec v} \over v^0}  {d^3{\vec v{\,'}} \over v{\,'}^0}\ \psi(v')^*\ \xi(v.v')\ \psi(v) \geq 0
\eeq
where $\psi(v)$ is an arbitrary function.

\subsection{Inconsistency with the sum rules of an exponential form for the Isgur-Wise function} \hspace*{\parindent}

We have seen that if the sum rules are satisfied for a $j = 0$ IW function $\xi(w)$ then, for any function $\psi(v)$, we have the positivity condition (\ref{114e}).\par
We will now show that, for an exponential form, a function that one could guess from the harmonic oscillator potential : 

\beq
\label{115e}
\xi(w) = exp\left [ -c(w-1) \right ]
\eeq
one can find a function $\psi(v)$ for which the integral in (\ref{114e}) is {\it strictly negative}, and this will prove that the exponential form of the IW function is incompatible with the sum rules.\par
For our purpose, it is enough to consider radial $\psi(v)$ functions :

\beq
\label{116e}
\psi(v) = \phi(|{\vec v}|)	
\eeq
First, let us integrate over the angles of ${\vec v}$ and $\vec v{\,'}$. Adopting just in a few lines below the notation $|{\vec v}| = v$ and $|{\vec v}{\,'}| = v'$, we obtain for the integral :

$$ \int {d^3{\vec v} \over v^0}  {d^3{\vec v{\,'}} \over v{\,'}^0}\ \phi(|\vec v{\,'}|)^*\ exp\left [-c((v.v')-1) \right ]\ \phi(|{\vec v}|) $$
$$ = 8\pi^2\ e^c \int_0^\infty {dv \over v^0} \int_0^\infty  {dv' \over v{\,'}^0} \int_{-1}^{1} ds\ \phi(v')^*\ exp[-c(v^0v'^0-vv's)]v^2 v'^2\ \phi(v) $$
$$ = 8\pi^2\ {e^c \over c} \int_0^\infty {dv \over v^0} \int_0^\infty  {dv' \over v{\,'}^0} \int_{-1}^{1} ds\ \phi(v')^*\phi(v) {d \over ds} \left ( exp[-c(v^0v'^0-vv's)] \right ) vv' $$
$$ = 8\pi^2{e^c \over c} \int_0^\infty {dv \over v^0} \int_0^\infty  {dv' \over v{\,'}^0}\ \phi(v')^*\phi(v) \left ( exp[-c(v^0v'^0-vv')] - exp[-c(v^0v'^0+vv')] \right ) vv' $$
Next we change the variables of integration :

\beq
\label{117e}
v = sh(\eta) \qquad\qquad\qquad v' = sh(\eta')
\eeq
and this gives :

$$ \int {d^3{\vec v} \over v^0}  {d^3{\vec v{\,'}} \over v{\,'}^0}\ \phi(|\vec v{\,'}|)^*\ exp\left [-c((v.v')-1) \right ]\ \phi(|{\vec v}|) $$
$$ = 8\pi^2\ {e^c \over c} \int_0^\infty \int_0^\infty d\eta\ d\eta'\ sh(\eta') \phi(sh(\eta'))^*\ sh(\eta)\phi(sh(\eta)) $$
$$ \left ( exp[-c\ ch(\eta'-\eta)] -  exp[-c\ ch(\eta'+\eta)] \right ) $$
$$ = 4\pi^2\ {e^c \over c} \int_{-\infty}^{\infty} \int_{-\infty}^{\infty} d\eta\ d\eta'\ sh(\eta') \phi(sh(\eta'))^*\ exp[-c\ ch(\eta'-\eta)]\ sh(\eta)\phi(sh(\eta)) $$
In the last line, the function $\phi(v)$ is extended to negative arguments by

\beq
\label{118e}
\phi(-v) = \phi(v)
\eeq

Introducing the functions $f(\eta)$ and $K(\eta)$ by :

\beq
\label{119e}
f(\eta) = sh(\eta)\ \phi(sh(\eta)) \qquad\qquad\qquad K(\eta) = exp[-c\ ch(\eta)]
\eeq
the last result writes

$$  \int {d^3{\vec v} \over v^0}  {d^3{\vec v{\,'}} \over v{\,'}^0}\ \phi(|\vec v{\,'}|)^*\ exp\left [-c((v.v')-1) \right ]\ \phi(|{\vec v}|) $$
\beq
\label{120e}
= 4\pi^2\ {e^c \over c} \int_{-\infty}^{\infty} \int_{-\infty}^{\infty} d\eta\ d\eta'\ f(\eta')^*\ K(\eta'-\eta)\ f(\eta)
\eeq
But we have here the matrix element of a convolution operator, and this is diagonalized by Fourier transforming. So, introducing

\beq
\label{121e}
\tilde{f}(\rho) = {1 \over 2\pi} \int_{-\infty}^{\infty} e^{i\rho \eta}\ f(\eta)\ d\eta
\eeq

\beq
\label{122e}
\tilde{K}(\rho) = \int_{-\infty}^{\infty} e^{i\rho \eta}\ K(\eta)\ d\eta
\eeq
we have

\beq
\label{123e}
\int_{-\infty}^{\infty} \int_{-\infty}^{\infty} d\eta\ d\eta'\ f(\eta')^*\ K(\eta'-\eta)\ f(\eta) = 2\pi \int_{-\infty}^{\infty} \tilde{K}(\rho)\ |\tilde{f}(\rho)|^2\ d\rho 
\eeq
And finally we obtain

\beq
\label{124bise}
\int {d^3{\vec v} \over v^0}  {d^3{\vec v{\,'}} \over v{\,'}^0}\ \phi(|\vec v{\,'}|)^*\ exp\left [-c((v.v')-1) \right ]\ \phi(|{\vec v}|)  = 8\pi^3\ {e^c \over c} \int_{-\infty}^{\infty} \tilde{K}(\rho)\ |\tilde{f}(\rho)|^2\ d\rho
\eeq
with $\tilde{f}(\rho)$ and $\tilde{K}(\rho)$ given by (\ref{118e}), (\ref{119e}), (\ref{121e}) and (\ref{122e}).\par
Now the Sommerfeld integral representation of the Macdonald function : 

\beq
\label{125bise}
K_{\nu}(z) = {1 \over 2} \int_{-\infty}^{\infty} exp[-z\ ch(t)]\ e^{\nu t}\ dt
\eeq
gives the following expression of $\tilde{K}(\rho)$ :

\beq
\label{126bise}
\tilde{K}(\rho) = 2\ K_{i\rho}(c)
\eeq
and we have :

\beq
\label{127bise}
\int {d^3{\vec v} \over v^0}  {d^3{\vec v{\,'}} \over v{\,'}^0}\ \phi(|\vec v{\,'}|)^*\ exp\left [-c((v.v')-1) \right ]\ \phi(|{\vec v}|) = 16\pi^3\ {e^c \over c} \int_{-\infty}^{\infty} K_{i\rho}(c)\ |\tilde{f}(\rho)|^2\ d\rho
\eeq

Then, we observe that, whatever $c > 0$, the function $\rho \to K_{i\rho}(c)$ {\it takes negative values}, as shown by the asymptotic formula

\beq
\label{128bise}
K_{i\rho}(c) \sim \sqrt{{2\pi \over \rho}}\ e^{-\rho \pi/2}\ cos\!\left[ \rho \left ( log\! \left ({2\rho \over c} \right ) - 1 \right ) - {\pi \over 4} \right] \qquad\qquad (\rho >> c) 
\eeq
So, taking for $\tilde{f}(\rho)$ a function peaked at a point $\rho_0$ where $K_{i\rho_0}(c) < 0$, the r.h.s. of (\ref{127bise}) will be $< 0$. Reversing the steps from $\psi(v)$ (where now $v$ is the quadrivector velocity) to $\tilde{f}(\rho)$, we then have a function $\psi(v)$ for which the integral in (\ref{114e}) is $< 0$. 

\section{Equivalence between the sum rule approach and the Lorentz group approach} \hspace*{\parindent}

We show here that the Lorentz group approach, introduced in the present work, is in fact equivalent to the (generalized Bjorken) sum rules. So, it must be considered just as a powerful way of exploring the consequences of these SR.\par
Here we must be more specific. When we say generalized Bjorken SR we do not mean the SR involving higher moments like Voloshin's \cite{11bisr} and its generalizations \cite{11terr}, but those that concern zero order moments. Of course, these sum rules are also not the ones of the QCD Sum Rule approach \`a la Shifman et al.\par 
For baryons we mean the SR that we have formulated in all generality in \cite{8r} and, for mesons, Uraltsev SR \cite{6r} and the SR formulated in \cite{7r}, that allow to obtain bounds on all the derivatives of the IW function at zero recoil.\par
That the Lorentz group approach implies the sum rules is already stressed in \cite{2r}. In fact, the SR are just completeness relations in the Hilbert space of the light components :

\beq
\label{127-1e}
< v_f, j', \epsilon', \alpha'|v_i, j, \epsilon, \alpha >
\eeq
$$ = \sum_{j'',M,\beta} < v_f, j', \epsilon', \alpha'|v, j'', \epsilon^M, \beta > < v, j'', \epsilon^M, \beta|v_i, j, \epsilon, \alpha > $$
when the overlaps are expressed in terms of the IW functions.\par
By the way, the fact that the $j = 0$ IW function $\xi(w)$ is of positive type, as expressed by (\ref{108e}), has also a very simple proof in the Lorentz group approach. Indeed, following (\ref{26bise}), let us write

\beq
\label{127-2e}
\xi(w) =\ < U(B_{v'})\psi_0|U(B_v)\psi_0>
\eeq
where $B_v$ is the boost transforming the rest velocity $v_0$ into $v$. Then we have :

$$ \sum_{i,j=1}^N a_i^*a_j\ \xi(v_i.v_j) = \sum_{i,j=1}^N a_i^*a_j < U(B_{v_i})\psi_0|U(B_{v_j})\psi_0> $$ 
\beq
\label{127-3e}
=\ < \sum_{i=1}^N a_i U(B_{v_i})\psi_0|\sum_{j=1}^N a_j U(B_{v_j})\psi_0>\ = \|{\sum_{j=1}^N a_j U(B_{v_j})\psi_0} \|^2 \geq 0
\eeq

Let us now turn to the proof that the sum rule approach implies the Lorentz group approach, so that the results in this work are in fact consequences of the sum rules. The proof is based on a theorem about functions of positive type on a group (see Dixmier \cite{10r}).\par
A function $f(\Lambda)$ on the group $SL(2,C)$ is of positive type when

\beq
\label{127-4e}
\sum_{i,j=1}^N a_i^*a_j\ f(\Lambda_i^{-1}\Lambda_j) \geq 0
\eeq
for any $N \geq 1$, any complex numbers $a_1,... ,a_N$, and any $\Lambda_1,... ,\Lambda_N \in SL(2,C)$. Then, according to theorem 13.4.5 in \cite{10r} ({\it C*-alg\`ebres}), for any such function $f(\Lambda)$ of positive type, there exists a unitary representation $U(\Lambda)$ of $SL(2,C)$ in a Hilbert space $\mathcal{H}$, and an element $\phi_0 \in \mathcal{H}$, such that

\beq
\label{127-5e}
f(\Lambda) =\ < \phi_0|U(\Lambda)\phi_0> 
\eeq

Moreover, one may assume that $\mathcal{H}$ contains no invariant strict subspace containing $\phi_0$, and then $\mathcal{H}$, $U(\Lambda)$ and $\phi_0$ are unique up to isomorphisms. This will not be used here, but it will show that a $j = 0$ IW function completely determines the Lorentz representation and the scalar state (or in fact the sub-representation generated by the scalar state).\par
Now, in Section 7 we have proved, by using only the sum rules, the positivity type property (\ref{108e}) of the $j = 0$ IW function $\xi(w)$. To apply the theorem quoted above, we must define, from $\xi(w)$, a function $f(\Lambda)$ of positive type on the group $SL(2,C)$. This is done as follows :

\beq
\label{127-6e}
f(\Lambda) = \xi( (\Lambda v_0)^0 ) \qquad \qquad \qquad v_0 = (1,0,0,0)  
\eeq
Indeed we have

\beq
\label{127-7e}
f(\Lambda_i^{-1}\Lambda_j) = \xi(v_0.\Lambda_i^{-1}\Lambda_j v_0) = \xi(\Lambda_iv_0.\Lambda_j v_0) 
\eeq
and, taking $v_i = \Lambda_iv_0$ and $v_j = \Lambda_jv_0$ in (\ref{108e}), one sees that (\ref{127-4e}) is satisfied.\par
We may conclude, from the sum rules, and from the positivity type property of $\xi(w)$ which follows from them, that there exists a unitary representation $U(\Lambda)$ of $SL(2,C)$ in a Hilbert space $\mathcal{H}$, and an element $\phi_0 \in \mathcal{H}$, such that 

\beq
\label{127-8e}
\xi( (\Lambda v_0)^0 ) =\ < \phi_0|U(\Lambda)\phi_0>
\eeq
But this is just the expression (\ref{26bise}) of $\xi(w)$ which ocurred in the Lorentz group approach.\par
There is a last point to be proved, namely that the state $\phi_0 \in \mathcal{H}$ which occurs here is a scalar under rotations. This results from the following property of $f(\Lambda)$ as defined by (\ref{127-6e}) :

\beq
\label{127-9e}
f(R\Lambda) = f(\Lambda) \qquad \qquad \qquad R \in SU(2)
\eeq
Looking at (\ref{127-5e}), this implies

\beq
\label{127-10e}
< U(R)\phi_0-\phi_0|U(\Lambda)\phi_0> \ = 0
\eeq
for any $R \in SU(2)$ and any $\Lambda \in SL(2,C)$. Applying this with $\Lambda = 1$ and $\Lambda = R$, one obtains

\beq
\label{127-11e}
\|U(R)\phi_0-\phi_0\|^2 =\ < U(R)\phi_0-\phi_0|U(R)\phi_0-\phi_0> \ = 0 
\eeq
so that

\beq
\label{127-12e}
U(R)\phi_0 = \phi_0 
\eeq
as we wanted to demonstrate.

\section{Another application of the Lorentz group approach : consistency test for any ansatz of the Isgur-Wise function} \hspace*{\parindent}

Section 7 was based only on the traditional method of (generalized Bjorken) sum rules, leaving aside for a while the representation of the Lorentz group and its decomposition into irreducible representations. We have shown that the sum rules imply a positive type property of the IW function, and that, in the case of an exponential IW function, there are functions on which the associated quadratic form is in fact negative.\par
Here we return to the main method of this work. As a result, we obtain a systematic way of testing the consistency of any ansatz for the IW function. We prove again (in a more comprehensive way) the inconsistency of the exponential form, but, as a positive result, we propose other possible forms for the IW function that we demonstrate to be consistent.\par

\subsection{Inversion of the integral representation of the Isgur-Wise function} \hspace*{\parindent}

Now, the fundamental question to be solved is the inversion of the integral representation (\ref{62e}), namely, given a function $\xi(w)$, to find the measures $d\nu_p(\rho)$, $d\nu_s(\rho)$, and a number $\nu_t$ such that (\ref{62e}) holds.\par
The solution to that problem will be found by Fourier transforming the function $sh(\tau)\ \xi(ch(\tau))$. After multiplying by $sh(\tau)$, the dependence on $\tau$ of the r.h.s. of (\ref{62e}) appears via the functions $sin(\rho\tau)$ and $sh(\rho\tau)$ :

\beq
\label{124e}
sh(\tau)\ \xi(ch(\tau)) = \int_{[0,\infty[}	{sin(\rho\tau) \over \rho}\ d\nu_p(\rho) + \int_{]0,1[} {sh(\rho\tau) \over \rho}\ d\nu_s(\rho) + \nu_t\ sh(\tau)
\eeq

Defining 

\beq
\label{124bis1e}
\widehat{\xi} (\tau) = sh(\tau)\ \xi(ch(\tau))
\eeq

a rather symbolic calculation of the Fourier transform gives :

\beq
\label{125e}
(\mathcal{F}\widehat{\xi})(\sigma) = {1 \over 2\pi} \int_{-\infty}^{\infty} e^{i\tau\sigma} sh(\tau)\  \xi(ch(\tau))\ d\tau 
\eeq
$$ = {i \over 2}  \int_{[0,\infty[} {1 \over \rho} \left [ \delta(\sigma-\rho)-\delta(\sigma+\rho) \right ] d\nu_p(\rho) + {1 \over 2}  \int_{]0,1[} {1 \over \rho} \left [ \delta(\sigma-i\rho)-\delta(\sigma+i\rho) \right ] d\nu_s(\rho) $$	
$$ +\ {1 \over 2}  \left [ \delta(\sigma-i)-\delta(\sigma+i) \right ] \nu_t $$
Here, the mathematical oriented reader is probably horrified, because a Dirac function with complex argument is not even a distribution ! The usual (L. Schwartz) Fourier transformation of a distribution apply to a subclass, the tempered distributions, and gives tempered distributions. However, fulfilling our need, there is a theory of Fourier transformation of {\it all} distributions by Gelfand and Chilov \cite{13r}. The point is that, generally, a distribution is a linear functional on the space  $\mathcal{D}(R)$ of the smooth functions with bounded support. The Fourier transform of a distribution will then be a linear functional on the space $\mathcal{Z}(R)$ made up of the Fourier transforms of the functions in $\mathcal{D}(R)$. This space is fully described in \cite{13r}. Due to the bounded support of the functions in  $\mathcal{D}(R)$, the functions in $\mathcal{Z}(R)$ extend to entire functions (analytic in the whole complex plane), and, for example, the distribution $\delta(\sigma-i\rho)$ on  $\mathcal{Z}(R)$ is defined by :

$$ \int_{-\infty}^{\infty} \delta(\sigma-i\rho)\ f(\sigma)\ d\sigma = f(i\rho) \qquad\qquad\qquad f \in \mathcal{Z}(R) $$  

The precise definition (as a linear functional on $\mathcal{Z}(R)$) of the Fourier transform in (\ref{125e}) is then as follows : 

\beq
\label{126e}
\left (\mathcal{F}\widehat{\xi},\tilde{u} \right ) = \int_{-\infty}^{\infty} sh(\tau)\ \xi(ch(\tau))\ u(\tau)\ d\tau 
\eeq
with 

$$ \tilde{u}(\rho) = \int_{-\infty}^{\infty} e^{-i\tau\rho}\ u(\tau)\ d\tau \qquad\qquad\qquad u \in \mathcal{D}(R) $$

The calculation of the r.h.s. of (\ref{126e}) using (\ref{124e}) involves justified exchanges of integrals, and is almost trivial, giving  

\beq
\label{127e}
\left (\mathcal{F}\widehat{\xi},\tilde{u} \right ) = {i \over 2} \int_{[0,\infty[}	{1 \over \rho} \left [ \tilde{u}(\rho)-\tilde{u}(-\rho)\right ] d\nu_p(\rho)
\eeq 
$$ + {1 \over 2} \int_{]0,1[} {1 \over \rho} \left [ \tilde{u}(i\rho)-\tilde{u}(-i\rho)\right ] d\nu_s(\rho) 
+ {1 \over 2} \left [ \tilde{u}(i)-\tilde{u}(-i)\right ] \nu_t $$ 
for any $\tilde{u} \in \mathcal{Z}(R)$. The above discussion shows that this condition (\ref{127e}) on $\xi(w)$ is fully equivalent to (\ref{62e}). 

\subsection{Illustration by the exponential form} \hspace*{\parindent}

Let us see what this gives for an exponential form $\xi(w) = e^{-c(w-1)}$ of the IW function. The function $sh(\tau)\ \xi(ch(\tau))$ is integrable in this case, and its Fourier transform is an ordinary function, given by a convergent integral, which is easily calculated :

$$ (\mathcal{F}\widehat{\xi})(\rho) = {1 \over 2\pi}\ e^c \int_{-\infty}^{\infty} e^{i\tau\rho}\  sh(\tau)\  exp[{-c\ ch(\tau)}]\ d\tau $$   
$$ = - {1 \over 2\pi} {e^c \over c} \int_{-\infty}^{\infty} e^{i\tau\rho}\ {d \over d\tau}\ exp[{-c\ ch(\tau)}]\ d\tau $$
\beq
\label{129e}
= {i \over 2\pi} {e^c \over c} \rho \int_{-\infty}^{\infty} e^{i\tau\rho}\ exp[{-c\ ch(\tau)}]\ d\tau = {i \over \pi} {e^c \over c} \rho\ K_{i\rho}(c)
\eeq
So, we have

\beq
\label{130e}
\left (\mathcal{F}\widehat{\xi}, \tilde{u} \right ) = {i \over \pi} {e^c \over c} \int_{-\infty}^{\infty} \rho\ K_{i\rho}(c)\  \tilde{u}(\rho)\ d\rho  
\eeq
This is indeed of the form (\ref{127e}) (taking into account $K_{-i\rho}(c) = K_{i\rho}(c)$), with the following measures

\beq
\label{131e}
d\nu_p(\rho) = {2 \over \pi} {e^c \over c} \rho^2 K_{i\rho}(c)\ d\rho \qquad\qquad d\nu_s(\rho) = 0 \qquad\qquad \nu_t = 0  
\eeq   
and for the integral representation (\ref{62e}), this gives 

\beq
\label{132e}
exp[-c(w-1)] = {2 \over \pi} {e^c \over c} \int_0^{\infty} \rho^2\ K_{i\rho}(c)\ {sin(\rho\tau) \over {\rho\ sh(\tau)}}\ d\rho \qquad\qquad (w = ch(\tau))
\eeq 
But (for any value $c > 0$), the function $K_{i\rho}(c)$ of $\rho$ takes {\it negative values}, as we have seen from (\ref{128bise}), so that the measure $d\nu_p(\rho)$ in (\ref{131e}) is not a positive one, contrarily to what is required in (\ref{62e}).\par
Nevertheless, we cannot yet conclude at the inconsistency of the exponential form. After all, as far as we know, a clever choice of the positive measures in (\ref{62e}) could perhaps give the same result as (\ref{132e}). What is still needed is a {\it unicity result}, namely, we must show that, if a function $\xi(w)$ can be put in the form (\ref{62e}), then the measures $d\nu_p(\rho)$, $d\nu_s(\rho)$ and the number $\nu_t$ are unique. Also, we must show this without assuming these measures and this number positive (in order for instance to apply this result also to (\ref{132e})).

\subsection{Unicity of the representation of the Isgur-Wise function} \hspace*{\parindent}

Let us now demonstrate that the measures $d\nu_p(\rho)$, $d\nu_s(\rho)$ and the number $\nu_t$ are unique in the representation (\ref{62e}) for the IW function. Moreover, here $d\nu_p(\rho)$, $d\nu_s(\rho)$ and $\nu_t$ are not assumed positive. \par
To be precise, let us specify that we consider here only bounded measures. A measure $d\nu$ is said to be bounded when $\int|d\nu| < \infty$. Then the integrals of bounded functions, as in (\ref{62e}), are convergent. The positive measures in (\ref{62e}) are bounded due to the condition (\ref{63e}).\par
Actually, an elementary unicity proof is obtained if one takes the {\it Laplace transform} of $sh(\tau)\ \xi(ch(\tau))$. The integral giving the Laplace transform of the r.h.s. of (\ref{124e}) is convergent for $Im(z) > 1$. So the Laplace transform of $sh(\tau)\ \xi(ch(\tau))$ is an analytic function in the half-plane $Im(z) > 1$, and a simple calculation, valid for $Im(z) > 1$, gives :

$$ \int_{[0,\infty[} sh(\tau)\ \xi(ch(\tau))\ e^{iz\tau}\ d\tau $$
\beq
\label{133e}
= \int_{[0,\infty[}	{d\nu_p(\rho) \over \rho^2-z^2}  + \int_{]0,1[} {d\nu_s(\rho) \over -\rho^2-z^2} + {\nu_t \over -1-z^2} = \int_{[-1,\infty[}	{d\nu(\sigma) \over \sigma-z^2} 
\eeq 
In the last line we have introduced a measure $d\nu(\sigma)$ on the set $[-1,\infty[$ whose restrictions to the subsets ${\{-1\}}$, $]-1,0[$ and $[0,\infty[$ are related to $\nu_t$, $d\nu_s(\rho)$ and $d\nu_p(\rho)$ by simple changes of variable, so that it is the same thing to know $d\nu(\sigma)$ or to know $\nu_t$, $d\nu_s(\rho)$ and $d\nu_p(\rho)$.\par
Replacing $z$ by $\sqrt{z}$ with $Im(\sqrt{z}) > 1$ in (\ref{133e}), we have a function $f_{\xi}(z)$, depending only on the function $\xi(w)$, given by

\beq
\label{134e}
f_{\xi}(z) = \int_0^{\infty}\ sh(\tau)\ \xi(ch(\tau))\ e^{i\tau\sqrt{z}}\ d\tau = \int_{[-1,\infty[}{d\nu(\sigma) \over \sigma-z} 
\eeq 
Independently of $\nu_t$, $d\nu_s(\rho)$ and $d\nu_p(\rho)$, this function $f_{\xi}(z)$ is analytic when $Im(\sqrt{z}) > 1$ (for one of the two determinations of $\sqrt{z}$). This domain in $z$ excludes a parabolic region containing $[-1,\infty[$ but, according to the last member of (\ref{134e}), $f_{\xi}(z)$ has an analytic continuation in the whole complex plane cut by $[-1,\infty[$. And according to the analytic functions theory, such an analytic continuation is unique.\par
Now we are done because, as well known, the measure $d\nu(\sigma)$ can be recovered from the discontinuity across the cut. Precisely, one has

\beq
\label{135e}
\int h(\sigma)\ d\nu(\sigma) = lim_{\epsilon \to 0}\ {1 \over 2\pi i} \int \left [f_{\xi}(\sigma+i\epsilon)-f_{\xi}(\sigma-i\epsilon) \right ] h(\sigma)\ d{\sigma}
\eeq 
for any continuous function $h(\sigma)$ going to zero at $\infty$, and this defines $d\nu(\sigma)$.\par
This ends the present proof (in the Lorentz group approach) of the inconsistency of the exponential form for the IW function.

\subsection{Summary of the general method to test the consistency of any ansatz for the Isgur-Wise function} \hspace*{\parindent}

With Subsections 9.1 and 9.3, we have now a general method to test the consistency of any form for the IW function in the case $j = 0$, that we summarize now.\par
Namely, given an ansatz for the function $\xi(w)$ :

- Compute the Fourier transform of $sh(\tau)\ \xi(ch(\tau))$ (possibly in the Guelfand-Chilov generalized sense \cite{13r}), as defined by (\ref{126e}).

- When this Fourier transform cannot be written in the form (\ref{127e}), the ansatz is inconsistent.

- When this Fourier transform is written in the form (\ref{127e}), the ansatz is consistent if the measures $d\nu_p(\rho)$, $d\nu_s(\rho)$ and the number $\nu_t$ are positive, and inconsistent if not.

\subsection{Ansatze for the Isgur-Wise function compatible with the sum rules} \hspace*{\parindent}

\par \vskip 3 truemm

\underline{Example 1}

\par \vskip 3 truemm

We now use this method to establish the {\it consistency} of

\beq
\label{136e}
\xi(w) = \left ( {2 \over 1+w} \right )^{2c}
\eeq 
{\it {for any slope $c \geq {1 \over 4}$}} (and inconsistency for $0 < c < {1 \over 4}$).\par
One has

\beq
\label{137e}
\widehat{\xi}(\tau) = sh(\tau)\ \xi(ch(\tau)) = 2\ sh(\tau/2)\ ch(\tau/2)^{-4c+1}
\eeq
When $c > {1 \over 2}$, this function is integrable, and its Fourier transform is an ordinary function, given by a convergent integral, which can be calculated :

\beq
\label{138e}
(\mathcal{F}\widehat{\xi})(\rho) = {1 \over \pi} \int_{-\infty}^{\infty}\ sh(\tau/2)\ ch(\tau/2)^{-4c+1}\ e^{i\tau\rho}\ d\tau
\eeq 
$$ = - {1 \over {2c-1}} {1 \over \pi} \int_{-\infty}^{\infty}\ \left [ {d \over d\tau} ch(\tau/2)^{-4c+2} \right ] e^{i\tau\rho} d\tau = {1 \over {2c-1}}\ i\rho\ {1 \over \pi} \int_{-\infty}^{\infty}\ ch(\tau/2)^{-4c+2}  e^{i\tau\rho} d\tau $$
and by the change of variable $t = e^\tau$ we have :

$$ (\mathcal{F}\widehat{\xi})(\rho) = 4^{2c-1} {1 \over {2c-1}}\ i\rho\ {1 \over 2\pi} \int_{0}^{\infty}\ (1+t)^{-4c+2}\ t^{i\rho+2c-2}\ dt $$
By the change of variable $t = {s \over {1-s}}$ we obtain :
$$ (\mathcal{F}\widehat{\xi})(\rho) = 4^{2c-1} {1 \over {2c-1}}\ i\rho\ {1 \over 2\pi} \int_{0}^{1}\ s^{i\rho+2c-2}\ (1-s)^{-i\rho+2c-2}\ ds $$
$$ = 4^{2c} i\rho {1 \over 2\pi} {\Gamma(2c+i\rho-1)\Gamma(2c-i\rho-1) \over \Gamma(4c-1)} $$
So, (\ref{126e}) writes, for the ansatz (\ref{136e}) :

\beq
\label{139e}
\left (\mathcal{F}\widehat{\xi},\tilde{u} \right ) = i {4^{2c} \over 2\pi} \int_{-\infty}^{\infty}\ \rho\  {\Gamma(2c+i\rho-1)\Gamma(2c-i\rho-1) \over \Gamma(4c-1)}\ \tilde{u}(\rho) \ d\rho
\eeq 
This is indeed of the form (\ref{127e}) with the following measures

\beq
\label{140e}
d\nu_p(\rho) = {4^{2c} \over \pi} \rho^2\  {|\Gamma(2c+i\rho-1)|^2 \over \Gamma(4c-1)} d\rho \qquad\qquad d\nu_s(\rho) = 0 \qquad\qquad \nu_t = 0
\eeq
Since these measures are positive, the consistency of (\ref{136e}) is established for the slopes $c > {1 \over 2}$.\par
To treat the cases $c \leq {1 \over 2}$, we have to find the generalized Fourier transform of $sh(\tau)\ \xi(ch(\tau)$. To that end, we use analytic continuation in $c$. The above calculation of the Fourier transform (\ref{139e}) is valid for complex $c$ provided $Re(c) > {1 \over 2}$. Considering (\ref{126e}), eq. (\ref{139e}) writes

$$ \int_{-\infty}^{\infty} sh(\tau)\ \left ( {2 \over 1+w} \right )^{2c}\ u(\tau) $$ 
\beq
\label{141e}
=  i\ {4^{2c} \over 2\pi} \int_{-\infty}^{\infty}\ \rho\  {\Gamma(2c+i\rho-1)\Gamma(2c-i\rho-1) \over \Gamma(4c-1)}\ \tilde{u}(\rho) \ d\rho
\eeq
for any $u \in \mathcal{D}(R)$. This formula is proved for $Re(c) > {1 \over 2}$, but the left hand side is an entire function of $c$ (due to the bounded support of $u$). So, the r.h.s. must have an analytic continuation, which will give the needed generalized Fourier transform of $\widehat {\xi}(\tau)$.\par
First, we can directly take the limit $c \to {1 \over 2}$ in (\ref{141e}), and also in (\ref{139e}) and (\ref{140e}), establishing the consistency of (\ref{136e}) also for the slope $c = {1 \over 2}$.\par
It is more tricky to go to $Re(c) < {1 \over 2}$. Let us write

\beq
\label{142e}
c = {1 \over 2}\ (1+\gamma e^{i\theta}) 
\eeq
with $\gamma > 0$ small, and see what happens when $\theta$ goes from $0$ to $\pi$. Writing 

\beq
\label{143e}
\Gamma(2c+i\rho-1)\Gamma(2c-i\rho-1) = {\Gamma(\gamma e^{i\theta}+i\rho+1)\Gamma(\gamma e^{i\theta}-i\rho+1) \over (\rho-i\gamma e^{i\theta})(\rho+i\gamma e^{i\theta})}
\eeq
one sees that, in the r.h.s. of (\ref{143e}), there are two poles, $\rho = \pm i\gamma e^{i\theta}$, which, when $\theta$ goes to ${\pi \over 2}$ from below, approach the path of integration (which is the real axis). To have an analytic function, one must then deform the path of integration to avoid these poles. For ${\pi \over 2} < \theta < {3\pi \over 2}$ the new path can be decomposed into the old path (the real axis), a counterclockwise small circle around the pole $\rho = i\gamma e^{i\theta}$ in the lower complex half-plane, and a clockwise small circle around the pole $\rho = - i\gamma e^{i\theta}$ in the upper half plane. The integrals along the small circles are given by the residue theorem, and the obtained analytic continuation of eq. (\ref{141e}) to $0 < Re(c) < {1 \over 2}$ is :

$$ \int_{-\infty}^{\infty} sh(\tau)\left ( {2 \over 1+w} \right )^{2c} u(\tau)\ d\tau $$ 
$$ =  i\ {4^{2c} \over 2\pi} \int_{-\infty}^{\infty}\ \rho\  {\Gamma(2c+i\rho-1)\Gamma(2c-i\rho-1) \over \Gamma(4c-1)}\ \tilde{u}(\rho)\ d\rho $$ 
\beq
\label{144e} 
+\ 2^{4c-1} [\tilde{u}(i(1-2c))-\tilde{u}(-i(1-2c))]
\eeq
When $c$ is real with $0 < c < {1 \over 2}$, this is indeed of the form (\ref{127e}) with the following measures

$$ d\nu_p(\rho) = {4^{2c} \over \pi} \rho^2\  {|\Gamma(2c+i\rho-1)|^2 \over \Gamma(4c-1)} d\rho $$
$$ d\nu_s(\rho) = (1-2c)\ 2^{4c}\ \delta(\rho-(1-2c)) $$ 
\beq
\label{145e} 
\nu_t = 0 \qquad\qquad\qquad\qquad\qquad\qquad
\eeq
Since these measures are positive when ${1 \over 4} \leq c$, the consistency of (\ref{136e}) is established for the slopes ${1 \over 4} \leq c < {1 \over 2}$. If $0 < c < {1 \over 4}$, the measure $d \nu_p(\rho)$ is {\it negative} and inconsistency follows.\par
The decomposition (\ref{62e}) of the ansatz (\ref{136e}) into irreducible IW functions (given by (\ref{59e}), (\ref{60e}), (\ref{61e})) is : 
  
$$ \left ( {2 \over 1+w} \right )^{2c} = {4^{2c} \over \pi} \int_0^\infty \rho^2\  {|\Gamma(2c+i\rho-1)|^2 \over \Gamma(4c-1)}\ \xi_{p,0,\rho}(w)\ d\rho $$ 
\beq
\label{146e} 
+\ \theta(1-2c)\ (1-2c)\ 2^{4c}\ \xi_{s,1-2c}(w) 
\eeq

When $c \geq {1 \over 2}$, the decomposition of the representation of the Lorentz group involves only the irreducible representations of the principal series, with a continuous combination of all ($n = 0$) of them. When ${1 \over 4} < c < {1 \over 2}$, we have a direct sum of such a continuous combination and of {\it one} of the irreducible representations of the supplementary series. When $c = {1 \over 4}$ the representation is reduced to the $\rho = {1 \over 2}$ irreducible representation of the supplementary series :

\beq
\label{147e} 
\left ( {2 \over 1+w} \right )^{1/2} = \xi_{s,1/2}(w)  
\eeq

When $0 < c < {1 \over 4}$, the weight of $\xi_{p,0,\rho}$ is $< 0$, and this of course cannot occur from decomposition of a representation.\par

\par \vskip 3 truemm

\underline{Example 2}

\par \vskip 3 truemm

As another application, we establish the consistency of the IW function

\beq
\label{147bis1e} 
\xi(w) = {1 \over \left [ {1 + {c \over 2} (w-1)} \right ]^{2}}  
\eeq
{\it for any slope $c \geq 1$} (and inconsistency for $0 < c < 1$).\par

The function

\beq
\label{147bis2e} 
\widehat{\xi}(\tau) = sh(\tau)\ \xi(ch(\tau)) = {sh(\tau) \over \left [ {1 + {c \over 2} (ch(\tau)-1)} \right ]^{2}}  
\eeq
is integrable (for $c > 0$), and its Fourier transform is an ordinary function, given by a convergent integral :

\beq
\label{147bis3e}
(\mathcal{F}\widehat{\xi})(\rho) = {1 \over {2\pi}} \int_{-\infty}^{\infty}\ {sh(\tau) \over \left [ {1 + {c \over 2} (ch(\tau)-1)} \right ]^{2}}\ e^{i\tau\rho}\ d\tau
\eeq

Let us compute it. One has 

$$ (\mathcal{F}\widehat{\xi})(\rho) = -{2 \over c} {1 \over {2\pi}} \int_{-\infty}^{\infty}\ \left [{d \over d{\tau}} {1 \over {1 + {c \over 2} (ch(\tau)-1)}} \right ]\ e^{i\tau\rho}\ d\tau $$
$$ = {2 \over c}\ i\rho\ {1 \over {2\pi}} \int_{0}^{\infty}{1 \over {1 + {c \over 2} (ch(\tau)-1)}}\ e^{i\tau\rho}\ d\tau \qquad \qquad (t = e^{\tau})$$
$$ = {2 \over c}\ i\rho\ {1 \over {2\pi}} \int_{0}^{\infty}{ t^{i\rho} \over {t + {c \over 4} (1-t)^2}}\ dt \qquad \qquad\qquad \qquad $$
The denominator is written as follows :
$$ t + {c \over 4}\ (1-t)^2 = {c \over 4}\ (t-t_1)(t-t_2) \qquad \qquad t_{1,2} = 1 - {2 \over c} \pm \sqrt{\left(1- {2 \over c}\right )^2 - 1} $$
and the integral is obtained by a calculus of residues : 

$$ (\mathcal{F}\widehat{\xi})(\rho) = {8 \over c^2}\ i\rho\ {1 \over {2\pi}} \int_{0}^{\infty}{ t^{i\rho} \over (t -t_1)(t -t_2)}\ dt $$
$$ = {4 \over c^2}\ {i\rho \over sh(\pi\rho)}\ {1 \over {2\pi}} \int_{0}^{\infty}{{(-t-i0)^{i\rho}-(-t+i0)^{i\rho}} \over (t -t_1)(t -t_2)}\ dt $$
$$ = -\ {4 \over c^2}\ {\rho \over sh(\pi\rho)} \left (res|_{t=t_1} + res|_{t=t_2} \right ) {(-t)^{i\rho} \over (t -t_1)(t -t_2)} $$
$$ = -\ {4 \over c^2}\ {\rho \over sh(\pi\rho)} {{(-t_1)^{i\rho}-(-t_2)^{i\rho}} \over t_1 -t_2} $$
So, (\ref{126e}) writes

\beq
\label{147bis4e}
\left (\mathcal{F}\widehat{\xi},\tilde{u} \right ) = -\ {4 \over c^2}\ \int_{-\infty}^{\infty}{\rho \over sh(\pi\rho)} {{(-t_1)^{i\rho}-(-t_2)^{i\rho}} \over t_1 -t_2}\ \tilde{u}(\rho)\ d\rho
\eeq
This is indeed of the form (\ref{127e}) with the following measures

\beq
\label{147bis5e}
d\nu_p(\rho) = i\ {8 \over c^2} {\rho^2 \over sh(\pi\rho)} {{(-t_1)^{i\rho}-(-t_2)^{i\rho}} \over t_1 -t_2}\ d\rho, \qquad d\nu_s(\rho) = 0, \qquad \nu_t = 0 
\eeq
and the decomposition (\ref{62e}) of (\ref{147bis1e}) into irreducible IW functions (\ref{59e})-(\ref{61e}) is :

\beq
\label{147bis6e}
{1 \over \left [ {1 + {c \over 2} (w-1)} \right ]^{2}} = i\ {8 \over c^2} \int_{0}^{\infty} {\rho^2 \over sh(\pi\rho)} {{(-t_1)^{i\rho}-(-t_2)^{i\rho}} \over t_1 -t_2}\ \xi_{p,0,\rho}(w)\ d\rho 
\eeq

Now, when $0 < c < 1$, one has $1 < {2 \over c} - 1$ and, replacing $c$ by a parameter $\gamma > 0$ by $ch(\gamma) = {2 \over c} - 1$, one has $t_1 = -e^{-\gamma}$ and $t_1 = -e^{\gamma}$. Then (\ref{147bis6e}) writes :

\beq
\label{147bis7e}
{1 \over \left [ {1 + {c \over 2} (w-1)} \right ]^{2}} = \ {8 \over c^2} \int_{0}^{\infty} {\rho^2 \over sh(\pi\rho)} {sin(\gamma\rho) \over sh(\gamma)}\ \xi_{p,0,\rho}(w)\ d\rho 
\eeq
and when $1 < c$, one has $-1 < {2 \over c} - 1 < 1$ and, replacing c by a parameter $0 < \gamma < \pi$ by $cos(\gamma) = {2 \over c} - 1$, one has $t_1 = -e^{i\gamma}$ and $t_1 = -e^{-i\gamma}$. Then (\ref{147bis6e}) writes :

\beq
\label{147bis8e}
{1 \over \left [ {1 + {c \over 2} (w-1)} \right ]^{2}} = \ {8 \over c^2} \int_{0}^{\infty} {\rho^2 \over sh(\pi\rho)} {sh(\gamma\rho) \over sh(\gamma)}\ \xi_{p,0,\rho}(w)\ d\rho 
\eeq
For the case $c = 1$, we can take the limit $\gamma \to 0$ in (\ref{147bis7e}) or (\ref{147bis8e}).

\beq
\label{147bis9e}
{1 \over \left [ {1 + {c \over 2} (w-1)} \right ]^{2}} = \ {8 \over c^2} \int_{0}^{\infty} {\rho^3 \over sh(\pi\rho)} \ \xi_{p,0,\rho}(w)\ d\rho 
\eeq

Since the weight of $\xi_{p,0,\rho}(w)$ in (\ref{147bis7e}) takes negative values, the ansatz  (\ref{147bis1e}) is incompatible with the sum rules when $0 < c < 1$.\par
Since the weight of $\xi_{p,0,\rho}(w)$ in (\ref{147bis8e}) and (\ref{147bis9e}) is positive, the ansatz (\ref{147bis1e}) is compatible with the sum rules when $1 \leq c$.

\par \vskip 3 truemm

\underline{Example 3}

\par \vskip 3 truemm

As a last application, we establish the consistency with the sum rules of

\beq
\label{148bis1e}
\xi(w) = e^{-c\tau} = {1 \over (w+\sqrt{w^2-1})^c}
\eeq
for any value $c\geq0$ of the parameter. Here $c$ is not the slope. In fact, for $c > 0$ all the derivatives at $w = 1$ are infinite, so that the bounds on these derivatives are helpless in this case.\par

We have to compute the Fourier transform of the function

\beq
\label{148bis2e}
\widehat{\xi}(\tau) = sh(\tau)\ \xi(ch(\tau)) = sh(\tau)\ e^{-c|\tau|}
\eeq
When $Re(c) > 1$, this functions is integrable, and its Fourier transform is an ordinary function given by a convergent integral :

\beq
\label{148bis3e}
(\mathcal{F}\widehat{\xi})(\rho) = {1 \over {2\pi}} \int^{\infty}_{-\infty} sh(\tau)\ e^{-c|\tau|}\ e^{i\tau\rho}\ d\tau
\eeq
which is easily calculated :

\beq
\label{148bis4e}
(\mathcal{F}\widehat{\xi})(\rho) = {1 \over {2\pi}}\ i\rho\ {4c \over [(c-1)^2+\rho^2][(c+1)^2+\rho^2]}
\eeq
This is indeed of the form (\ref{127e}) with the following measures

\beq
\label{148bis5e}
d\nu_p(\rho) = {4c \over {2\pi}}\ {\rho^2 \over [(c-1)^2+\rho^2][(c+1)^2+\rho^2]}\ d\rho, \ \ \ d\nu_s(\rho) = 0, \ \ \ \nu_t = 0
\eeq
Since these measures are positive for $c$ real, the consistency of (\ref{148bis1e}) is established for $c > 1$.\par

To treat the cases $c \leq 1$, we have to find the generalized Fourier transform of $\widehat{\xi}(\tau)$, and to that end, as in the case of the "dipolar" form (\ref{136e}), we use analytic continuation in $c$. Combining the general definition (\ref{126e}) of the Fourier transform with the Fourier transform (\ref{148bis2e}) already calculated, we have : 

\beq
\label{148bis6e}
\int^{\infty}_{-\infty}\widehat{\xi}(\tau)\ u(\tau)\ d\tau = i\ {2c \over \pi}\int^{\infty}_{-\infty} {\rho \over [(c-1)^2+\rho^2][(c+1)^2+\rho^2]}\ \tilde{u}(\rho)\ d\rho
\eeq
for any $u \in \mathcal{D}(R)$. This formula is proved for $Re(c) > 1$, but the left hand side is an entire function of $c$ (due to the bounded support of $u$), so the r.h.s. must have an analytic continuation, which will give the needed generalized Fourier transform of $\widehat{\xi}(\tau)$.\par

First, we can directly take the limit $c \to 1$ in (\ref{148bis6e}), and also in (\ref{148bis4e}) (in the sense of tempered distributions) and (\ref{148bis5e}), establishing the consistency of (\ref{148bis1e}) also for $c = 1$.\par

Going down to $Re(c) < 1$ is quite similar to the case of the "dipolar" form (\ref{136e}). In the integrand of (\ref{148bis6e}), there are two poles

\beq
\label{148bis7e}
\rho = \pm\ i\ (c-1)
\eeq
which, when $c$ goes around $c = 1$ from above, approach the path of integration (which is the real axis). To have an analytic function, one must then deform the path of integration to avoid these poles. The new path can be decomposed into the old path (the real axis), a counterclockwise small circle around the pole $\rho = i(c-1)$ in the lower complex half-plane, and a clockwise small circle around the $\rho = - i(c-1)$ in the upper complex half-plane. The integrals along the small circles are given by the residue theorem, and the obtained analytic continuation of (\ref{148bis2e}) to $0 \leq Re(c) < 1$ is :

\beq
\label{148bis8e}
\int^{\infty}_{-\infty}\widehat{\xi}(\tau)\ u(\tau)\ d\tau = i\ {2c \over \pi}\int^{\infty}_{-\infty} {\rho \over [(c-1)^2+\rho^2][(c+1)^2+\rho^2]}\ \tilde{u}(\rho)\ d\rho
\eeq
 $\qquad \qquad \qquad \qquad \qquad \qquad +\ {1 \over 2}\ [\tilde{u}(i(1-c)) - \tilde{u}(-i(1-c))] $\par
 When $c$ is real with $0 < c < 1$, this is indeed of the form (\ref{127e}) with the following measures
 
\beq
\label{148bis9e}
d\nu_p(\rho) = {4c \over {2\pi}}\ {\rho^2 \over [(c-1)^2+\rho^2][(c+1)^2+\rho^2]}\ d\rho
\eeq
$ \qquad \qquad  \ d\nu_s(\rho) = (1-c)\ \delta(\rho-(1-c))\ d\rho \qquad \qquad \qquad \nu_t = 0 $\par

Since these measures are positive when $0 < c < 1$, the consistency of (\ref{148bis1e}) is established for these values of $c$. For $c = 0$, $\xi(w) = 1$ is just the irreducible IW function given by the trivial representation. So, we have consistency for all $c \geq 0$.\par
The decomposition (\ref{62e}) of (\ref{148bis1e}) into irreducible IW functions (given by (\ref{59e}), (\ref{60e}) and (\ref{61e})) is, for $c > 0$ :

\beq
\label{148bis10e}
{1 \over (w+\sqrt{w^2-1})^c} =  {4c \over {2\pi}} \int_{0}^{\infty} {\rho^2 \over [(c-1)^2+\rho^2][(c+1)^2+\rho^2]}\ \xi_{p,0,\rho}(w)\ d\rho
\eeq
$ \qquad \qquad \qquad \qquad \qquad \ \ +\ \theta(1-c)\ (1-c)\ \xi_{s,1-c}(w) $\par

The fact that all the derivatives at $w = 1$ are infinite is due to the slow decrease of the measure $d\nu_p(\rho)$, for which the moments $\mu_k =\ < x^k >$ defined in (\ref{76bise}) are divergent for $k \geq 1$.

\section{Phenomenological applications} \hspace*{\parindent}

Before concluding, let us summarize the main phenomenological consequences of the present paper. From a practical perspective, we have a number of interesting results for possible simple forms of the $j = 0$ IW function.\par
We have illustrated our different general results with some one-parameter ansatze for the IW function (from now on we make the replacement $c = \rho_\Lambda^2$ for the slope, when it is finite), namely :\par 

(i) The "dipole" form (\ref{104bise}) :
 
\beq
\label{149e}
\xi(w) = \left ( 2 \over {w+1} \right )^{2\rho_\Lambda^2} \qquad \qquad \qquad \rho_\Lambda^2 \geq {1 \over 4}
\eeq

(ii) The true dipole shape (\ref{105bis1e}) :

\beq
\label{149bise}
\xi(w) = {1 \over \left[{1+{\rho_\Lambda^2 \over 2}(w-1)}\right]^2} \qquad \qquad \qquad \rho_\Lambda^2 \geq 1
\eeq

(iii) The form found in Subsection 6.2 (\ref{104e}) :

\beq
\label{149bis1e}
\xi(w) = {sh\left (\tau \sqrt{1-3\rho_\Lambda^2} \right ) \over sh(\tau) \sqrt{1 - 3\rho_\Lambda^2}} = {sin\left (\tau \sqrt{3\rho_\Lambda^2-1} \right ) \over sh(\tau) \sqrt{3\rho_\Lambda^2-1}} \qquad\qquad w = ch(\tau) \qquad \rho_\Lambda^2 \geq 0
\eeq

(iv) The form proposed in Subsection 9.5

\beq
\label{149bis2e}
\xi(w) = {1 \over {(w+\sqrt{w^2-1}})^c}  \qquad \qquad c \geq 0
\eeq
for which at $w = 1$ {\it all derivatives are infinite} if $c > 1$. 

Let us comment on these different possible one-parameter models for the IW function and briefly remind the results obtained above.
\par \vskip 3 truemm
(i) The "dipole" form (\ref{149e}) was proposed in the case of the {\it meson} IW function \cite{13bisr}. For this function we have shown in Subsection 6.1 that all the bounds 
(\ref{97e})-(\ref{99e}) imply $ \rho_\Lambda^2 \geq {1 \over 4} $, suggesting that this form is acceptable if this lower bound is fulfilled. Indeed, following the consistency test of Section 9, we have shown that this form is consistent for any slope $\rho_\Lambda^2 \geq {1 \over 4}$ (and inconsistent for $0 < \rho_\Lambda^2 < {1 \over 4}$). 
\par \vskip 3 truemm
(ii) The true dipole form is a model proposed in \cite{14r} for baryon decay (\ref{149bise}). For it, we have shown in Subsection 6.1 that the bounds 
(\ref{97e})-(\ref{99e}) imply $\rho_\Lambda^2 \geq {2 \over 3}$, $ \rho_\Lambda^2 \geq {10 \over 13} $, $ \rho_\Lambda^2 \geq 0.86 $, lower bounds that slowly converge towards 1 with the constraints on incresing order derivatives. Indeed, following the consistency test of Section 9, we have shown that this form is consistent for any slope $\rho_\Lambda^2 \geq 1$ (and inconsistent for $0 < \rho_\Lambda^2 < 1$). 
\par \vskip 3 truemm
(iii) The form (\ref{149bis1e}) is a result of the present paper if the lower bound on the curvature (\ref{97e}) is saturated, i.e. $\sigma_\Lambda^2 = {3 \over 5} \rho_\Lambda^2 (1 + \rho_\Lambda^2)$. It satisfies all the constraints for any value of the slope $\rho_\Lambda^2 \geq 0$. 
\par \vskip 3 truemm
(iv) The form (\ref{149bis2e}) is interesting because it satisfies all the constraints for $c \geq 0$, with all its derivatives being infinite at zero recoil if $c > 0$.\par
\par \vskip 3 truemm
These simple one-parameter forms will be useful in the future to fit the differential decay width for the process $\Lambda_b \to \Lambda_c \ell \overline{\nu}_{\ell}$ with quite different possibilities, and thus guess a possible variation of $|V_{cb}|$. In this sense, the form (\ref{149bis2e}) constitutes an extreme case, since all the derivatives are infinite at $w = 1$.
\par \vskip 3 truemm
Another extreme case is the one-parameter form 
\beq
\label{149bis3e}
\xi(w) = 1 - e^{-c/(w-1)}  \qquad \qquad c > 0
\eeq
for which $\xi(w) \to 0$ for $w \to \infty$, and all its derivatives vanish at $w = 1$. But according to Subsection 6.2 (see (\ref{103e})), if the slope vanishes one gets $\xi(w) = 1$, and only the trivial representation contributes to the integral formula (\ref{62e}), with $\nu_t = 1$. Therefore, the ansatz (\ref{149bis3e}) is inconsistent with the sum rules. 
 
\par \vskip 3 truemm

Finally, let us comment on the exponential form (\ref{102bise}),

\beq
\label{148e}
\xi(w) = exp[-\rho_\Lambda^2(w-1)] \qquad \qquad \qquad \ \ \ 
\eeq
that was proposed by E. Jenkins et al. \cite{14bisr} and by M. Pervin et al. \cite{14r}. In the large $N_c$ and heavy quark limit studied in \cite{14bisr} there is an important subtlety that we discuss at the end of this Section.\par 
For the exponential form we have shown in Subsection 6.1 that the bounds 
(\ref{97e})-(\ref{99e}) imply respectively $ \rho_\Lambda^2 \geq 1.5 $, $ \rho_\Lambda^2 \geq 2.5 $, $ \rho_\Lambda^2 \geq 4.28 $. The lower bound on the slope grows with the constraints on higher and higher derivatives, suggesting that the exponential form is not consistent. Indeed, we have demonstrated in Section 7 from the sum rules that the IW function is of {\it positive type}, and that the exponential ansatz is inconsistent with this property for any value of the slope $\rho_\Lambda^2 > 0$. We have exposed an alternative demonstration following the general consistency test formulated for any form of the IW function in Subsection 9.2.
\par \vskip 3 truemm

Let us make a final remark on the exponential form (\ref{148e}). This form was suggested in the paper by E. Jenkins et al. \cite{14bisr} within a model based on QCD in the heavy quark and large $N_c$ limits, with a slope of the order :

\beq
\label{151e}
\rho_\Lambda^2 = \lambda N_c^{3/2} \qquad\qquad\qquad\qquad \lambda = O(1)
\eeq 
However, one must keep in mind that formula (\ref{148e}) with the slope (\ref{151e}) is valid, according to \cite{14bisr}, for 

\beq
\label{152e}
w-1 = O(N_c^{-3/2})
\eeq
This means that in this scheme (\ref{148e}) would be valid in the heavy quark limit, but for {\it fixed} 

\beq
\label{153e}
x = \rho_\Lambda^2\ (w-1)
\eeq
As we said in \cite{8r}, the bound (\ref{97e}), obtained in the physical situation $N_c = 3$ is trivially satisfied in the large $N_c$ limit, as it is obvious from (\ref{148e}) and (\ref{151e}). However, the phenomenological guess (3.8) from \cite{14bisr}, $ \rho_\Lambda^2 = 1.3 $, slightly violates the bound, and we have more generally demonstrated that the exponential form is inconsistent. \par
But there is a subtle point concerning the exponential form (\ref{148e}) at {\it fixed} $x = \rho_\Lambda^2(w-1)$  (\ref{153e}). Indeed, the "dipole" form (\ref{149e}), that satisfies all the theoretical constraints, becomes, performing the change of variables (\ref{153e}) and taking the limit $\rho_\Lambda^2 = O(N_c^{3/2}) \to \infty$ :

\beq
\label{154e}
\xi(w) = \left ( 2 \over {w+1} \right )^{2\rho_\Lambda^2} = \left ( 1 \over {1 + {x \over {2\rho_\Lambda^2}}} \right )^{2\rho_\Lambda^2} = e^{-2\rho_\Lambda^2 Log\left (1 + {x \over {2\rho_\Lambda^2}} \right )} \sim e^{-x} = e^{-\rho_\Lambda^2(w-1)}
\eeq
Therefore, within the conditions (\ref{151e}) and (\ref{152e}), i.e. for a very large slope and an infinitesimally small phase space, the exponential form (\ref{148e}) can be rigorously replaced by the "dipole" form (\ref{149e}), that satisfies all the theoretical constraints formulated in the present paper. Therefore, for finite slope and the whole phase space it would be convenient on theoretical grounds to replace the exponential form by the "dipole" form.\par

\section{Conclusion} \hspace*{\parindent}

The present paper explores new methods to study Isgur-Wise functions based on the Lorentz group. The IW function is expressed in terms of the scalar product of the initial and final light clouds of the heavy hadron, that involves a unitary representation of the Lorentz group. The method uses the decomposition of this unitary representation into irreducible representations and under the SU(2) sub-group of rotations. The approach has practical consequences, namely constraints on the possible IW functions, that can be applied to the different parametrizations proposed in the literature.\par
For the moment, we have applied this method to the case of a light cloud with $j^P = 0^+$, relevant to the decay $\Lambda_b \to \Lambda_c \ell \overline{\nu}_{\ell}$. This case is more involved from the point of view of group theory than the ground state meson case  $j^P = {1 \over 2}^-$. We leave the latter, being more complicated from the spin point of view, for future work.\par
We have shown, in the present baryon case, that the enumeration and explicit formulae for the relevant irreducible representations allows to give an integral formula for the IW function $\xi(w)$ involving {\it positive measures}. Not only the {\it principal series} of the unitary representations of the Lorentz group appear, but also the so-called {\it supplementary series}. This powerful formula allows in turn to express the derivatives of the IW function at zero recoil as moments of a variable with positive values.\par The corresponding positivity constraints on determinants of these moments imply in turn bounds on the $k$-th derivative of the IW function $\xi^{(k)}(1)$ in terms of the lower derivatives $\xi^{(n)}(1)\ (n =0,1,...k-1)$. We have illustrated these bounds for three one-parameter models of the IW function proposed in the literature, namely the exponential form and two different kinds of the "dipole" forms. The exponential form, unlike the "dipole" forms, appears to be somewhat pathological in this respect.\par 
We have also demonstrated that if one of the bounds is saturated (e.g. the bound on the curvature in terms of the slope), then one gets a completely explicit and simple one-parameter form of the IW function.\par
Then we have used the sum rule approach \cite{8r}, and demonstrated that the IW function is {\it a function of positive type}. This allows to show, as an example, that the exponential form for the IW function is not consistent with this property.\par
We demonstrate also, using the positive type property, that the Lorentz group method developed in the present paper, and this is important, is equivalent to the sum rule approach. Moreover, the Lorentz group method sheds another light on the long distance physics, and summarizes all the possible constraints of the sum rule approach.\par
Finally, we have formulated a general consistency test for any given ansatz of the IW function. We have applied this criterium to several phenomenological one-parameter forms proposed in the literature, like the exponential and the "dipole" forms, and shown that the former is inconsistent, while the two latter forms are consistent when the slope satisfies some lower bounds.\par
Hopefully, LHCb will provide new data on the decay $\Lambda_b \to \Lambda_c \ell \overline{\nu}_{\ell}$, that we know has a large branching ratio, of the order of $5.10^{-2}$, measured at LEP,  roughly half of the total semileptonic rate $\Lambda_b \to X_c \ell \overline{\nu}_{\ell}$. One expects at LHCb roughly $3.10^{10}$ $\Lambda_b \to \Lambda_c \ell \overline{\nu}_{\ell}$ events/year \cite{15r} and possibly one could give a precise measurement of the differential rate. This measurement has a two-fold interest. One concerns heavy quark hadronic physics, namely in particular the shape of the IW function, the object of the present paper. The other is an independent useful exclusive determination of $|V_{cb}|$, since there is still at present some tension between the exclusive and the inclusive determinations in $B$ decays, the former giving a smaller value, although with a larger error.\par
For the decay $\Lambda_b \to \Lambda_c \ell \overline{\nu}_{\ell}$, theoretical work remains to be done. One should include radiative corrections within HQET and $1/m_Q$ corrections, as well as the Wilson coefficients that make the matching with the physical form factors, a program that was realized in the case of mesons by M. Dorsten \cite{16r}. This would allow to compare with the future data and with other theoretical or phenomenological schemes of baryon form factors at finite mass. Also, once these necessary improvements are realized, any future fit to the differential distribution of $\Lambda_b \to \Lambda_c \ell \overline{\nu}_{\ell}$ should take into account the constraints formulated here for the IW function.\par
It is important to apply the method of the present paper to mesons. In this case one has the complication of spin, since the light cloud has $j^P = {1 \over 2}^-$, but we have noticed that from the point of view of the Lorentz group the problem seems simpler because only the principal series of the representations of the Lorentz group appears. This program will be the object of a forthcoming work.

\par \vskip 6 truemm

\begin{center} 
{\large \bf Acknowledgements}\par
\end{center}

This work supported in part by the EU Contract No. MRTN-CT-2006-035482, ÒFLAVIAnetÓ.

\begin{appendix}

\section{Scalar products in Hilbert spaces of the supplementary series} 
\hspace*{\parindent} 

In this appendix, we describe a trick (which can be found in \cite{11r}) useful to compute scalar products 
(\ref{38e}) in the Hilbert space of a representation of the supplementary series.\par

The matrices $R \in SU(2)$ are of the form

$$R = \left( \begin{array}{cc} a & b \\ -\overline{b} & \overline{a} \end{array} \right) \qquad\qquad\qquad |a|^2+|b|^2 = 1\qquad\qquad\qquad\qquad (A.1)$$
Parametrising $a$ and $b$ by

$$a = cos\left( {\theta \over 2} \right) e^{i\phi} \qquad b = sin\left( {\theta \over 2} \right) e^{i\psi} \qquad (0 \leq \phi, \psi \leq 2\pi, 0 \leq \theta \leq \pi) \qquad (A.2)$$
the normalized invariant measure on $SU(2)$ writes

$$dR = {1 \over {8\pi^2}} sin\theta\ d\theta\ d\phi\ d\psi \qquad\qquad\qquad\qquad (A.3)$$

Defining $R(z,\alpha) \in SU(2)$ for $(z,\alpha) \in C \times [0,2\pi[$ by :

$$a = {1 \over \sqrt{1+|z|^2}}\ e^{i\alpha} \qquad\qquad\qquad b = {z \over \sqrt{1+|z|^2}}\ e^{-i\alpha} \qquad\qquad\qquad (A.4)$$
the Jacobian is :

$$dR(z,\alpha) = {1 \over {2\pi^2}} {1 \over \left (1+|z|^2 \right )^2}\ d^2z\ d\alpha \qquad\qquad\qquad\qquad (A.5)$$

Noting that, abreviating $R(z,\alpha)$ to $R$ and $R(z',\alpha')$ to $R'$, one has 

$$z = {R_{12} \over R_{22}} \qquad\qquad\qquad\qquad\qquad (A.6)$$
and (using $R_{12}'^{-1} = - R_{12}'$ and $R_{11}'^{-1} = R_{22}'$, which follow from $detR = 1$)

$$z' - z = {R_{12}' \over R_{22}'} - {R_{12} \over R_{22}} = {R_{12}'R_{22}-R_{22}'R_{12} \over R_{22}'R_{22}} = - {(R'^{-1}R)_{12} \over  R_{22}' R_{22}} \qquad\qquad (A.7)$$  
the scalar product (\ref{38e}), rewritten here for convenience

$$<\phi'|\phi> = \int \overline{\phi'(z')}\ |z'-z|^{2\rho-2}\ \phi(z)\ d^2z'\ d^2z \qquad\qquad (A.8)$$
can be written as follows

$$<\phi'|\phi> = \pi^2 \int |R_{22}'|^{-2\rho-2}\ \overline{\phi}'\!\left ({R_{12}' \over R_{22}'} \right) |(R'^{-1}R)_{12}|^{2\rho-2}\ |R_{22}|^{-2\rho-2}\ \phi\! \left ({R_{12} \over R_{22}} \right) dR' dR \qquad (A.9)$$
and by a change of varible $R \to R'R$, one obtains :

$$<\phi'|\phi> = \pi^2 \int |R_{22}'|^{-2\rho-2}\ \overline{\phi}'\! \left ({R_{12}' \over R_{22}'} \right) |R_{12}|^{2\rho-2}\ |(R'R)_{22}|^{-2\rho-2}\ \phi\!\left ({(R'R)_{12} \over (R'R)_{22}} \right) dR' dR \qquad (A.10)$$

Next, applying the transformation law (\ref{35e}) with $\Lambda = R'^{-1}$, we have

$$\left (U_{s,\rho}(R'^{-1}) \phi \right )\!(z) = |R_{11}'^{-1}-R_{21}'^{-1}z|^{-2\rho-2}\ \phi\! \left ({R_{22}'^{-1}z-R_{12}'^{-1} \over R_{11}'^{-1}-R_{21}'^{-1}z} \right) \qquad\qquad\qquad\qquad $$
$$\left (U_{s,\rho}(R'^{-1}) \phi \right )\!(z) = |R_{22}'+R_{21}'z|^{-2\rho-2}\ \phi\! \left ({R_{11}'z+R_{12}' \over R_{22}'+R_{21}'z} \right) \qquad\qquad\qquad\qquad\qquad $$
$$\left (U_{s,\rho}(R'^{-1}) \phi \right )\! \left ({R_{12} \over R_{22}} \right ) = \left ({{|R_{22}'R_{22}+R_{21}'R_{12}|} \over |R_{22}|} \right )^{-2\rho-2}\phi\! \left ({{R_{11}'R_{12}+R_{12}'R_{22}} \over {R_{22}'R_{22}+R_{21}'R_{12}}} \right ) \qquad$$
$$\left (U_{s,\rho}(R'^{-1})\phi \right )\! \left ({R_{12} \over R_{22}} \right ) = |R_{22}|^{2\rho+2}\  |(R'R)_{22}|^{-2\rho-2}\ \phi\! \left ({(R'R)_{12} \over (R'R)_{22}} \right ) \qquad \qquad (A.11)$$
This expresses a part of the integrand in (A.10), which becomes

$$ <\phi'|\phi>\ = \pi^2 \int |R_{22}'|^{-2\rho-2}\ \overline{\phi}'\!\left ({R_{12}' \over R_{22}'} \right) |R_{12}|^{2\rho-2} \qquad \qquad \qquad \qquad $$ 
$$ |R_{22}|^{-2\rho-2} \left (U_{s,\rho}(R'^{-1}) \phi \right )\!\left ({R_{12} \over R_{22}} \right ) dR' dR \qquad \qquad \qquad \qquad (A.12) $$
Returning to the variables $z'$ and $\alpha'$ for $R'$ and $z$ and $\alpha$ for $R$, we obtain the following expression for the scalar product (\ref{38e}) in the Hilbert space $\mathcal{H}_{s,\rho}$ of the representation in the supplementary series labelled by $\rho$ :

$$<\phi'|\phi>\ = {1 \over 2\pi} \int (1+|z'|^2)^{\rho-1}\ \overline{\phi'(z')}\ |z|^{2\rho-2} \left (U_{s,\rho}(R(z',\alpha')^{-1}) \phi \right )\! (z)\ d^2z\ d^2z'\ d\alpha' \qquad (A.13)$$
This result will be used in Appendix B and in Appendix C.

\par \vskip 1 truecm

\section{Calculation of normalization constants} 
\hspace*{\parindent} 

In this appendix, we compute the normalization constants in (\ref{45e}) and (\ref{47e}). 

\underline{Principal series}. 

Using the notation $R(z,\alpha)$ for the rotation defined by (A.4), we write (\ref{45e}) as follows :

$$\phi^{p,n,\rho}_{j,M}(z) = N^{p,n,\rho}_{j,M}\ (1+|z|^2)^{i\rho-1} D^j_{n/2,M}\!\left ({R(z,0)^{-1}} \right ) \quad\quad\quad\quad\quad (B.1)$$
where $N^{p,n,\rho}_{j,M}$ is the normalization constant here to be found. We have to compute the following scalar product.

$$<\phi^{p,n,\rho}_{j',M'}|\phi^{p,n,\rho}_{j,M}>\ = N^{p,n,\rho}_{j',M'}\ N^{p,n,\rho}_{j,M} \quad\quad \quad\quad \quad\quad \quad\quad \quad\quad (B.2)$$
$$\int (1+|z|^2)^{-2}\ D^{j'}_{n/2,M'}\!\left ({R(z,0)^{-1}} \right )^{*} D^j_{n/2,M}\!\left ({R(z,0)^{-1}} \right ) d^2z $$ 
From (A.4), one sees that :

$$R(z,\alpha) = R(z,0) \left( \begin{array}{cc} e^{i\alpha} & 0 \\ 0 & e^{-i\alpha} \end{array} \right) \quad\quad\quad\quad (B.3)$$
and because $D^j_{n/2,M}(R)$, as defined by (\ref{40e}), is a matrix element with on the left an eigenstate of the $Oz$ component of the angular momentum, with eigenvalue $n/2$, one has

$$D^j_{n/2,M}\!\left ({R(z,\alpha)^{-1}} \right ) = e^{in\alpha} D^j_{n/2,M}\!\left ({R(z,0)^{-1}} \right ) \quad\quad\quad (B.4)$$
and we may rewrite (B.2) as follows

$$<\phi^{p,n,\rho}_{j',M'}|\phi^{p,n,\rho}_{j,M}>\ = N^{p,n,\rho}_{j',M'}\ N^{p,n,\rho}_{j,M} \quad\quad \quad\quad \quad\quad \quad\quad \quad\quad (B.5)$$
$${1 \over {2\pi}} \int (1+|z|^2)^{-2}\ D^{j'}_{n/2,M'}\!\left ({R(z,\alpha)^{-1}} \right )^{*} D^j_{n/2,M}\!\left ({R(z,\alpha)^{-1}} \right ) d^2z\ d\alpha$$
since the integrand does not in fact depend on $\alpha$. Using the Jacobian (A.5), this gives

$$<\phi^{p,n,\rho}_{j',M'}|\phi^{p,n,\rho}_{j,M}>\ = N^{p,n,\rho}_{j',M'}\ N^{p,n,\rho}_{j,M}\ 
\pi \int D^{j'}_{n/2,M'}\!\left ({R^{-1}} \right )^{*} D^j_{n/2,M}\!\left ({R^{-1}} \right ) dR  \quad\quad (B.6)$$
By the change of variable of integration $R \to R^{-1}$, which leaves invariant the measure $dR$, this reduces to the scalar product (\ref{44e}) of the rotation matrix elements, and one obtains 

$$<\phi^{p,n,\rho}_{j',M'}|\phi^{p,n,\rho}_{j,M}>\ = N^{p,n,\rho}_{j',M'}\ N^{p,n,\rho}_{j,M}\ 
{\pi \over {2j+1}}\ \delta_{j,j'}\ \delta_{M,M'} \quad\quad\quad (B.7)$$
So, the normalization constant is

$$ N^{p,n,\rho}_{j,M} = \sqrt{2j+1 \over \pi} \quad\quad\quad\quad\quad\quad\quad\quad\quad (B.8)$$

\underline{Supplementary series}. 

Using the notation $R(z,\alpha)$ for the rotation defined by (A.4), we write (\ref{47e}) as follows :

$$\phi^{s,\rho}_{j,M}(z) = N^{s,\rho}_{j,M}\
(1+|z|^2)^{-\rho-1}\ D^j_{0,M}\! \left( R(z,0)^{-1} \right) \quad\quad\quad (B.9)$$
where $N^{s,\rho}_{j,M}$ is the normalization constant to be found.

Using (A.13) for the scalar product in the supplementary series, we have

$$<\phi^{s,\rho}_{j',M'}|\phi^{s,\rho}_{j,M}> = {1 \over 2\pi} \int (1+|z'|^2)^{\rho-1}\ \phi^{s,\rho}_{j',M'}(z')^{*}\  |z|^{2\rho-2}$$ 
$$\left (U_{s,\rho}(R(z',\alpha')^{-1}) \phi^{s,\rho}_{j,M} \right )\! (z)\ d^2z\ d^2z'\ d\alpha' \quad\quad\quad\quad\quad (B.10)$$
Now, because the $\phi^{s,\rho}_{j,M}$ (for $-j \leq M \leq j$) are the standard basis of the representation $j$ of $SU(2)$, we have 

$$\left (U_{s,\rho}(R(z',\alpha')^{-1} \right )\!\phi^{s,\rho}_{j,M} = \sum_{M''} D^j_{M'',M}\!\left (R(z',\alpha')^{-1} \right )\phi^{s,\rho}_{j,M''} \quad\quad (B.11)$$
Using (B.11) and (B.9), the scalar product (B.10) writes 

$$<\phi^{s,\rho}_{j',M'}|\phi^{s,\rho}_{j,M}> =  \sum_{M''} N^{s,\rho}_{j',M'} N^{s,\rho}_{j,M''}\ {1 \over 2\pi} \int |z|^{2\rho-2}(1+|z|^2)^{-\rho-1} D^{j}_{0,M''} \left( R(z,0)^{-1} \right) d^2z$$ 
$$\int (1+|z'|^2)^{-2}\ D^{j'}_{0,M'}\!\left( R(z',\alpha')^{-1} \right)^{*} D^j_{M'',M}\! \left( R(z',\alpha')^{-1} \right )\ d^2z'\ d\alpha'  \quad\quad (B.12)$$
where we have also used the fact that $D^{j'}_{0,M'}\! \left( R(z',\alpha')^{-1} \right)$ does not depend on $\alpha'$ (see (B.4)). Using the Jacobian (A.5), the second integral is 

$$2\pi^2 \int D^{j'}_{0,M'}\! \left( R^{-1} \right)^{*} D^{j}_{M'',M}\! \left( R^{-1} \right) dR = {2\pi^2 \over 2j+1}\ \delta_{j,j'}\ \delta_{0,M''}\ \delta_{M,M'}  \quad\quad (B.13)$$
so that (B.12) reduces to

$$<\phi^{s,\rho}_{j',M'}|\phi^{s,\rho}_{j,M}> =  N^{s,\rho}_{j',M'} N^{s,\rho}_{j,M}\ {\pi \over 2j+1}\ \delta_{j,j'}\ \delta_{M,M'}  \quad\quad\quad\quad$$ 
$$\int |z|^{2\rho-2}(1+|z|^2)^{-\rho-1} D^{j}_{0,0}\! \left( R(z,0)^{-1} \right) d^2z  \quad\quad\quad\quad\quad (B.14)$$ 
Now, from (\ref{43e}), with $a = {1 \over \sqrt{1+|z|^2}}$ and $b = - {z \over \sqrt{1+|z|^2}}$, we have

$$ D^{j}_{0,0}\! \left( R(z,0)^{-1} \right) = (1+|z|^2)^{-j} \sum_k\ (-1)^k  \left( \begin{array}{c} j \\ k \end{array} \right) \left( \begin{array}{c} j \\ j-k \end{array} \right) |z|^{2k} \quad\quad\quad (B.15)$$
The calculation of the remaining integral then goes as follows  

$$\int |z|^{2\rho-2}\ (1+|z|^2)^{-\rho-1} D^{j}_{0,0}\! \left( R(z,0)^{-1} \right) d^2z$$ 
$$ = \sum_{0 \leq k \leq j}  (-1)^k  \left( \begin{array}{c} j \\ k \end{array} \right) \left( \begin{array}{c} j \\ j-k \end{array} \right) \int |z|^{2\rho+2k-2}\ (1+|z|^2)^{-\rho-j-1} d^2z $$
$$ = \pi \sum_{0 \leq k \leq j}  (-1)^k  \left( \begin{array}{c} j \\ k \end{array} \right) \left( \begin{array}{c} j \\ j-k \end{array} \right) \int_0^\infty x^{\rho+k-1}\ (1+x)^{-\rho-j-1} dx \qquad \left( x \to {y \over 1-y} \right )$$
$$ = \pi \sum_{0 \leq k \leq j}  (-1)^k  \left( \begin{array}{c} j \\ k \end{array} \right) \left( \begin{array}{c} j \\ j-k \end{array} \right) \int_0^1 y^{\rho+k-1}\ (1-y)^{j-k} dy $$
$$ = \pi \sum_{0 \leq k \leq j}  (-1)^k  \left( \begin{array}{c} j \\ k \end{array} \right) \left( \begin{array}{c} j \\ j-k \end{array} \right) {\Gamma(\rho+k)\Gamma(j-k+1) \over \Gamma(\rho+j+1)} $$
$$ = \pi\ {j!\ \Gamma(\rho) \over \Gamma(\rho+j+1)} \sum_{k}\  (-1)^k  \left( \begin{array}{c} j \\ k \end{array} \right) {\Gamma(\rho+k) \over k!\ \Gamma(\rho)}$$ 
$$ = \pi\ {j!\ \Gamma(\rho) \over \Gamma(\rho+j+1)} \sum_{k}\  (-1)^k  \left( \begin{array}{c} j \\ k \end{array} \right)  \left( \begin{array}{c} \rho+k-1 \\ k \end{array} \right)$$  $$ = \pi\ {j!\ \Gamma(\rho) \over \Gamma(\rho+j+1)} \sum_{k} \left( \begin{array}{c} j \\ j-k \end{array} \right)  \left( \begin{array}{c} -\rho \\ k \end{array} \right) = \pi\ {j!\ \Gamma(\rho) \over \Gamma(\rho+j+1)} \left( \begin{array}{c} j-\rho \\ j \end{array} \right) $$
and gives 

$$\int |z|^{2\rho-2}\ (1+|z|^2)^{-\rho-1} D^{j}_{0,0}\! \left( R(z,0)^{-1} \right) d^2z = \pi\ {\Gamma(j-\rho+1)\ \Gamma(\rho) \over \Gamma(j+\rho+1)\ \Gamma(1-\rho)} \quad\quad\quad (B.16) $$
With (B.14), this gives the final result for the scalar product 

$$<\phi^{s,\rho}_{j',M'}|\phi^{s,\rho}_{j,M}>\ =  N^{s,\rho}_{j',M'} N^{s,\rho}_{j,0}\ {\pi^2 \over 2j+1}\ {\Gamma(j-\rho+1)\ \Gamma(\rho) \over \Gamma(j+\rho+1)\ \Gamma(1-\rho)}\ \delta_{j,j'}\ \delta_{M,M'} \quad\quad\quad (B.17) $$
So, the normalization constant is

$$N^{s,\rho}_{j,M} = {\sqrt{2j+1} \over \pi}\ \sqrt{{\Gamma(j+\rho+1)\ \Gamma(1-\rho) \over \Gamma(j-\rho+1)\ \Gamma(\rho)}} \quad\quad\quad\quad\quad\quad (B.18) $$ 
\\
 
\section{Calculation of the irreducible Isgur-Wise functions for the $j = 0$ case}  \hspace*{\parindent} 

In this Appendix, we compute the IW functions (\ref{59e}) and (\ref{60e}) for the $j = 0$ state in the irreducible representations of $SL(2,C)$. These functions are in fact known, and can be found in \cite{11r} (with some changes of notation) under the name of "elementary spherical functions".

\underline{Principal series}. 

The integral (\ref{56e}) for $\xi_{p,0,\rho}(w)$ is directly computed. Integrating over the angle gives :

$$ \xi_{p,0,\rho}(w) = \int_0^\infty (1+x)^{-i\rho-1}(e^{\tau}+e^{-\tau}x)^{i\rho-1} dx \quad\quad\quad\quad\quad\quad (C.1) $$
and by the change of variable $x = {y \over 1-y}$, we obtain :

$$ \xi_{p,0,\rho}(w) = \int_0^1 \left[e^{\tau}(1-y)+e^{-\tau}y\right]^{i\rho-1} dy = {sin(\rho\tau) \over \rho\ sh(\tau)} \quad\quad\quad\quad (C.2)$$

\underline{Supplementary series}. 

The integral for $\xi_{s,\rho}(w)$ is more involved. In order to use (A.13), it is convenient to rewrite $\xi_{s,\rho}(w)$ in the form

$$ \xi_{s,\rho}(w) =\ < U_{s,\rho}(\Lambda_{-\tau})\phi_{0,0}^{s,\rho}|\phi_{0,0}^{s,\rho} > \quad\quad\quad\quad\quad\quad (C.3) $$
where we have used the unitarity of $U_{s,\rho}(\Lambda)$,

$$U_{s,\rho}(\Lambda)^{\dagger} = U_{s,\rho}(\Lambda)^{-1} = U_{s,\rho}(\Lambda^{-1}) \qquad\qquad \Lambda_{\tau}^{-1} = \Lambda_{-\tau} \qquad\qquad (C.4)$$
In (A.13), we have then the following simplification

$$ \left (U_{s,\rho}(R(z',\alpha')^{-1}) \phi_{0,0}^{s,\rho} \right )\!(z) = \phi_{0,0}^{s,\rho}(z)  \qquad\qquad\qquad\qquad (C.5) $$
due to the fact that $\phi_{0,0}^{s,\rho}$ is a scalar under the subgroup $SU(2)$ of $SL(2,C)$, and we obtain

$$ \xi_{s,\rho}(w) = \left [\int (1+|z'|^2)^{\rho-1} \left (U_{s,\rho}(\Lambda_{-\tau}) \phi_{0,0}^{s,\rho} \right )\!(z')^{*} d^2z' \right ] \left [\int |z|^{2\rho-2}\ \phi_{0,0}^{s,\rho}(z)\ d^2z \right ] \qquad (C.6) $$
The function $\phi_{0,0}^{s,\rho}$ is given by (\ref{50e}) and $U_{s,\rho}(\Lambda_{-\tau}) \phi_{0,0}^{s,\rho}$ is then given by (\ref{39e})

$$ \phi^{s,\rho}_{0,0}(z) = {\sqrt{\rho} \over \pi} (1+|z|^2)^{-\rho-1} \qquad \left ( U_{s,\rho}(\Lambda_{-\tau})\phi_{0,0}^{s,\rho} \right )\!(z) = {\sqrt{\rho} \over \pi} \left (e^{-\tau}+e^{\tau}|z|^2 \right )^{-\rho-1} \qquad (C.7)$$
The integrals in (C.6) are then directly computed :

$$ \int (1+|z'|^2)^{\rho-1} \left (e^{-\tau}+e^{\tau}|z'|^2 \right )^{-\rho-1} d^2z' = \pi \int_0^\infty (1+x)^{\rho-1} \left (e^{-\tau}+e^{\tau}x \right )^{-\rho-1} dx $$
$$ =  \pi \int_0^1 \left [e^{-\tau}(1-y)+e^{\tau}y \right ]^{-\rho-1} dy = {\pi \over \rho} {e^{\rho\tau}-e^{-\rho\tau} \over e^{\tau}-e^{-\tau}} = \pi {sh(\rho\tau) \over \rho\ sh(\tau)} \qquad\qquad (C.8) $$

$$ \int |z|^{2\rho-2}\ (1+|z|^2)^{-\rho-1}\ d^2z = \pi \int_0^\infty x^{\rho-1}\ (1+x)^{-\rho-1}\ dx = \pi \int_0^1 y^{\rho-1}\ dy = {\pi \over \rho} \qquad (C.9) $$
and we obtain :

$$ \xi_{s,\rho}(w) = {sh(\rho \tau) \over {\rho\ sh(\tau)}} \qquad\qquad\qquad\qquad\qquad (C.10) $$

\par \vskip 1 truecm

\section{Expansion in powers of $(w-1)$ of the irreducible $j = 0$ Isgur-Wise function}  \hspace*{\parindent}

In this appendix, we obtain the whole expansion in powers of $w-1$ of the IW function for the $j = 0$ state in the irreducible representations of $SL(2,C)$.\par
We work with the function $\xi_x(w)$ defined by (\ref{68e}) which, when $x \geq 0$, covers all the cases
(\ref{59e}), (\ref{60e}) and (\ref{61e}). It is not easy to obtain the $w-1$ expansion directly from (\ref{68e}) since in this formula the dependence in $w$ occurs through $\tau = Arcch(w)$.\par

We now obtain an integral representation (D.3) for $\xi_x(w)$ in which the dependence on $w$ is explicit and simple. To this end, let us compute the integral

$$ I_a(w) = \int_0^\infty {s^a \over (1+2ws+s^2)}\ ds = \int_0^\infty {s^a \over (s+e^{\tau})(s+e^{-\tau})}\ ds \qquad\qquad (D.1) $$
which is convergent for $-1 < Re(a) < 1$. A standard calculus of residues gives :
$$  I_a(w) = - {\pi \over sin(\pi a)} {1 \over 2i\pi} \int_0^\infty {(-z-i0)^a-(-z+i0)^a \over (z+e^{\tau})(z+e^{-\tau})}\ dz $$
$$ = - {\pi \over sin(\pi a)} \left (res_{z = - e^{\tau}} + res_{z = - e^{-\tau}} \right ) {(-z)^a \over (z+e^{\tau})(z+e^{-\tau})} $$
$$ = - {\pi \over sin(\pi a)} \left [ {e^{a\tau} \over (-e^{\tau}+ e^{-\tau})} + {e^{-a\tau} \over (-e^{-\tau}+ e^{\tau})} \right ] = {\pi \over sin(\pi a)} {sh(a\tau) \over sh(\tau)} \qquad (D.2)$$
Then for $\xi_x(w) = {sh(\tau\sqrt{1-x}) \over sh(\tau)\sqrt{1-x}}$ we have 

$$ \xi_x(w) = {sin(\pi \sqrt{1-x}) \over \pi \sqrt{1-x}} \int_0^\infty {s^{\sqrt{1-x}} \over 1+2ws+s^2}\ ds \qquad\qquad\qquad (D.3)$$ 
valid for $x > 0$.\par

We can now expand at $w = 1$ :  

$$ \xi_x(w) = {sin(\pi \sqrt{1-x}) \over \pi \sqrt{1-x}} \sum_{k \geq 0}\ (-1)^k\ 2^k\ (w-1)^k \int_0^\infty {s^{k+\sqrt{1-x}} \over (1+s)^{2k+2}}\ ds \qquad\qquad (D.4) $$

The integral is directly calculated :

$$ \int_0^\infty {s^{k+\sqrt{1-x}} \over (1+s)^{2k+2}}\ ds = \int_0^1 t^{k+\sqrt{1-x}}\ (1-t)^{k-\sqrt{1-x}}\ dt \qquad\qquad (D.5) $$
$$ = {\Gamma(k+\sqrt{1-x}+1)\ \Gamma(k-\sqrt{1-x}+1) \over (2k+1)!} = {\pi \sqrt{1-x} \over sin(\pi \sqrt{1-x})} {1 \over (2k+1)!} \prod_{i = 1}^k [(i^2-1)+x] $$
where we have used 

$$ \Gamma(\sqrt{1-x}+1)\ \Gamma(-\sqrt{1-x}+1) =  {\pi \sqrt{1-x} \over sin(\pi \sqrt{1-x})} \qquad\qquad (D.6) $$
and we obtain

$$ \xi_x(w) = \sum_{k \geq 0}\ (-1)^k\ 2^k\ {1 \over (2k+1)!}\ \prod^k_{i=1}\ (x+i^2-1)\ (w-1)^k \qquad\qquad (D.7) $$
This deduction of (D.7) is valid only when $x > 0$. Considering the case $x = 0$, we have

$$ \prod^k_{i=1}\ (i^2-1) = \delta_{k,0} $$
so that the formula (D.7) reduces to $\xi_0(w) = 1$, and is therefore true also in this case.

\section{Sum rule for the Isgur-Wise function in the $j = 0$ case}

In \cite{8r}, from the OPE and the non-forward amplitude, we have demonstrated the following sum rule for the $j = 0$ case :

$$\xi(w_{if}) = \sum_n \sum_{L\geq0} \tau_L^{(n)}(w_i)^*\tau_L^{(n)}(w_f) \sum_{0 \leq k \leq L/2} C_{L,k} (w_i^2-1)^k (w_f^2-1)^k  \bigg \{ (w_iw_f-w_{if})^{L-2k}$$ 
$$ - {2 \over {2L+1}} \bigg [ (L-2k)(w_i+1)(w_f+1)(w_iw_f-w_{if})^{L-2k-1}+2k(w_iw_f-w_{if})^{L-2k} \bigg ]$$
 $$ + {2 \over (2L+1)^2} \bigg [ (L-2k)(3+4k)(w_i+1)(w_f+1)(w_iw_f-w_{if})^{L-2k-1} \qquad\qquad\qquad$$
 $$+(L-2k)(L-2k-1)(w_i+1)(w_f+1)(w_iw_f+w_i+w_f-1-2w_{if})(w_iw_f-w_{if})^{L-2k-2}$$
 $$+4k^2(w_iw_f-w_{if})^{L-2k}\bigg ] \bigg \} \qquad\qquad\qquad\qquad\qquad\qquad\qquad\qquad\qquad\qquad (E.1) $$
where $w_i = v_i.v'$, $w_f = v_f.v'$, $w_{if} = v_i.v_f$ and $v_i$, $v_f$, $v'$ are the initial, final and intermediate state four-velocities in the sum rule, $\tau_L^{(n)}(w)$ are the IW functions for the transition $0^+ \to L^P$ with $P=(-1)^L$, and the coefficients $C_{L,k} $ are given by

$$ C_{L,k} = (-1)^k {(L!)^2 \over (2L)!} {(2L-2k)! \over k!(L-k)!(L-2k)!}	\qquad\qquad\qquad\qquad\qquad (E.2)$$
From this sum rule we have demonstrated the inequalities for the slope (\ref{96e}) \cite{9r} and for the curvature (\ref{97e}).\par
We have recently realized that the expression for this sum rule can enormously be simplified.\par 
First, let us group the terms in (E.1), that gives :  

$$\xi(w_{if}) = \sum_n \sum_{L\geq0} \tau_L^{(n)}(w_i)^*\tau_L^{(n)}(w_f)\  {1 \over (2L+1)^2}\qquad\qquad\qquad\qquad\qquad (E.3)$$ 
$$ \sum_{0 \leq k \leq L/2} C_{L,k} (w_i^2-1)^k (w_f^2-1)^k(w_iw_f-w_{if})^{L-2k-2}\qquad\qquad\qquad\qquad\qquad $$
$$ \bigg \{ \bigg [(2L+1)^2-4(2L+1)k+8k^2 \bigg ](w_iw_f-w_{if})^2 - 2(L-2k)(L-2k-1)(w_i^2-1)^k (w_f^2-1) \bigg \}$$

Next, using the expression for the Legendre polynomials :

$$ P_L(x) = {1 \over 2^L} \sum_k\ (-1)^k {(2L-2k)! \over k!(L-k)!(L-2k)!}\ x^{L-2k} \qquad\qquad\qquad (E.4) $$
one gets

$$ \sum_{0 \leq k \leq L/2} C_{L,k}\ x^{L-2k} = 2^L {(L!)^2 \over (2L)!} P_L(x) \qquad\qquad\qquad (E.5) $$

Defining now the variable $x_{if}$ as :

$$ x_{if} = {w_iw_f-w_{if} \over \sqrt{(w_i^2-1)(w_f^2-1)}} \qquad\qquad\qquad\qquad (E.6) $$
one obtains, from (E.3), the following expression for the sum rule :

$$ \xi(w_{if}) = \sum_n \sum_{L\geq0} \tau_L^{(n)}(w_i)^*\tau_L^{(n)}(w_f) {1 \over (2L+1)^2} (w_i^2-1)^{L/2} (w_f^2-1)^{L/2} \qquad\qquad\qquad (E.7)$$ 
$$ \sum_{0 \leq k \leq L/2} C_{L,k}\ x_{if}^{L-2k} \bigg [2L^2+6L-8k+1-2(L-2k)(L-2k-1)(1-x_{if}^2)x_{if}^{-2}\bigg ] \qquad\qquad\qquad $$
that can be written in terms of derivatives of Legendre polynomials :

$$ \xi(w_{if}) = \sum_n \sum_{L\geq0} \tau_L^{(n)}(w_i)^*\tau_L^{(n)}(w_f) {2^L \over (2L+1)^2} {(L!)^2 \over (2L)!} (w_i^2-1)^{L/2} (w_f^2-1)^{L/2} \qquad\qquad (E.8)$$ 
$$ \bigg [(2L^2+2L+1)P_L(x_{if})+4x_{if}P'_L(x_{if})-2(1-x_{if}^2)P''_L(x_{if}) \bigg ] \qquad\qquad\qquad $$
and from the differential equation satisfied by the Legendre polynomials 

$$ (1-x^2)P''_L(x) - 2xP'_L(x) + L(L+1)P_L(x) = 0 \qquad\qquad\qquad (E.9) $$
one gets

$$ \xi(w_{if}) = \sum_n \sum_{L\geq0} {2^L(L!)^2 \over (2L)!}
\tau_L^{(n)}(w_i)^*\tau_L^{(n)}(w_f) (w_i^2-1)^{L/2} (w_f^2-1)^{L/2} P_L(x_{if}) \ \ \ \ (E.10)$$
that gives finally the simple expression for the sum rule, used in Section 7 :

$$\xi(w_{if}) = \sum_n \sum_{L\geq0} \  \tau_L^{(n)}(w_i)^*\tau_L^{(n)}(w_f) \sum_{0 \leq k \leq L/2} C_{L,k}\ (w_i^2-1)^k (w_f^2-1)^k(w_iw_f-w_{if})^{L-2k}
\ \ (E.11)$$

\end{appendix}

\end{document}